\documentclass{aa}  

\usepackage{graphicx}
\usepackage{booktabs}
\usepackage{supertabular}
\usepackage[version=4]{mhchem}
\usepackage{ulem}
\usepackage{xcolor}
\usepackage{amsmath}
\usepackage{sidecap}
\usepackage{txfonts}
\begin{document}

   \title{Importance of source structure on complex organics emission}

   \subtitle{III. Effect of disks around massive protostars}

   \author{P. Nazari
          \inst{1}
          \and
          B. Tabone\inst{2}
          \and
          G. P. Rosotti\inst{1, 3, 4}
          }

   \institute{Leiden Observatory, Leiden University, P.O. Box 9513, 2300 RA Leiden, the Netherlands\\ 
        \email{nazari@strw.leidenuniv.nl}
         \and
         Universit\'e Paris-Saclay, CNRS, Institut d’Astrophysique Spatiale, 91405 Orsay, France
         \and
         School of Physics and Astronomy, University of Leicester, Leicester LE1 7RH, UK
         \and
         Dipartimento di Fisica `Aldo Pontremoli', Universit\`{a} degli Studi di Milano, via G. Celoria 16, I-20133 Milano, Italy
             }

   \date{Received XXX; accepted YYY}

 
  \abstract
   {The hot molecular core phase of massive star formation shows emission from complex organic molecules. However, these species are only detected toward a fraction of high-mass protostars. In particular, there is a ${\sim} 2$ orders of magnitude spread in methanol emission intensity from high-mass protostars.}
   {The goal of this work is to answer the question whether high-mass disks can explain the lack of methanol emission from some massive protostellar systems.}
   {We consider an envelope-only and an envelope-plus-disk model and use RADMC-3D to calculate the methanol emission. High and low millimeter (mm) opacity dust (representing large and small dust distributions) are considered for both models separately and the methanol abundance is parameterized. Viscous heating is included due to the high accretion rates of these objects in the disk.}
   {In contrast with low-mass protostars, the presence of a disk does not significantly affect the temperature structure and methanol emission. The shadowing effect of the disk is not as important for high-mass objects and the disk mid-plane is hot because of viscous heating, which is effective due to the high accretion rates. The methanol emission is lower for models with high mm opacity dust because of the dust attenuation blocking the emission in the envelope and hiding it in the disk through continuum over-subtraction, but the disk needs to be large for this to become effective. A minimum disk size of ${\sim}2000-2500$\,au is needed (at $L=10^4$\,L$_{\odot}$) with large mm opacity dust for a factor of about one order of magnitude drop in the methanol emission compared with the envelope-only models with low mm opacity dust. Consistent with observations of infrared absorption lines toward high-mass protostars, we find a vertical temperature inversion, i.e. higher temperatures in the disk mid-plane than the disk surface, at radii ${\lesssim} 50$\,au for the models with $L=10^4$\,L$_{\odot}$ and large mm opacity dust as long as the envelope mass is ${\gtrsim}550$\,M$_{\odot}$ ($\dot{M}=3.6\times 10^{-3}$\,M$_{\odot}$\,yr$^{-1}$).} 
   {The large observed scatter in methanol emission from massive protostars can be mostly explained toward lower luminosity objects (${\sim} 10^3$\,L$_{\odot}$) with the envelope-plus-disk models including low and high mm opacity dust. The methanol emission variation toward sources with high luminosities ($\gtrsim 10^4$\,L$_{\odot}$) cannot be explained by models with or without a disk with relatively high gas-phase abundance of methanol. However, the $L/M$ of these objects suggest that they could be associated with hypercompact/ultracompact HII regions. Therefore, the low methanol emission toward the high-luminosity sources can be explained by them hosting an HII region where methanol is absent.}

   \keywords{Astrochemistry --
                Stars: massive --
                Stars: protostars --
                ISM: abundances --
                (ISM:) HII regions --
                Radiative transfer
               }

   \maketitle
%

\section{Introduction}

Protostellar systems are the hottest and thus the richest phase of star formation in gaseous complex organic molecules sublimating from the ices (\citealt{Herbst2009}; \citealt{Caselli2012} \citealt{Jorgensen2020}; \citealt{vantHoff2020c}). These species are detected toward both low- and high-mass protostars (e.g., \citealt{Blake1987}; \citealt{Ewine1995}; \citealt{Schilke1997}; \citealt{Cazaux2003}; \citealt{Beltran2009}; \citealt{Belloche2013}; \citealt{Jorgensen2016}; \citealt{Rivilla2017}; \citealt{Bogelund2018}; \citealt{Martin-Domenech2019};  \citealt{vangelder2020}; \citealt{Taniguchi2020}; \citealt{Gorai2021}). Among these species methanol is the most abundant and well-studied species and it is known to mostly form on the surfaces of interstellar dust grains (\citealt{Watanabe2002}; \citealt{Fuchs2009}). In this work we focus on methanol as a representative of complex organic species.  

Although many high-mass protostars do show millimeter (mm) emission from methanol, there are many that do not. In particular, \cite{vanGelder2022} surveyed the methanol mass toward a large number of low- and high-mass protostars (for the low-mass sample also see \citealt{Yang2021PEACHES} and \citealt{Belloche2020}). Those observations were taken with the Atacama Large Millimeter/submillimeter Array (ALMA). They found a four orders of magnitude scatter in warm methanol mass. \cite{vanGelder2022} discuss various reasons for such scatter including the possible effect of dust optical depth (\citealt{Rivilla2017}; \citealt{Lopez2017}; \citealt{DeSimone2020}) and presence of a disk (\citealt{Persson2016}). \cite{Nazari2022} investigated the effect of a disk and optically thick dust on lowering mm emission from methanol toward low-mass protostars using radiative transfer modelling. They found that both a disk and optically thick dust are necessary to explain the lack of methanol emission at mm wavelengths in these objects. However, it is not yet clear whether disks can explain this lack of methanol emission toward massive protostars. 

The reason that a disk in low-mass protostars lowers the emission is that it generally decreases the temperature of the environment through disk shadowing and creating a cold mid-plane (e.g., \citealt{Murillo2015}). Therefore, methanol molecules will be mostly frozen out and unable to emit in mm wavelengths. Moreover, the presence of optically thick dust in the disk causes the continuum over-subtraction effect and decreases the line flux even further (\citealt{Nazari2022}). This effect, as explained in detail in \citealt{Nazari2022}, occurs when the methanol molecules are on top of the dust in the disk and in between the dusty disk and the observer. Therefore, dust does not block the methanol emission. In this scenario, if the continuum emission is approximately as strong as the methanol emission, it will hide the methanol emission and continuum subtraction will produce an error.

High-mass protostars have much higher accretion rates than low-mass protostars (${\sim} 10^{-4}-10^{-3}$\,M$_{\odot}$\,yr$^{-1}$; \citealt{Hosokawa2009}; \citealt{Beuther2017}). This means that viscous heating in the disk mid-plane becomes important, especially for accretion rates above ${\sim}10^{-5}$\,M$_{\odot}$\,yr$^{-1}$ (\citealt{Harsono2015}). Observational evidence of such heating is the presence of mid-infrared absorption lines toward high-mass disks (\citealt{Knez2009}; \citealt{Barr2020}). This was interpreted as the colder disk surface absorbing the emission from the hotter gas in the mid-plane. Therefore, high-mass protostellar disks may not affect the methanol emission in a manner similar to the low-mass protostellar disks. 

Another complication of studying massive protostellar disks is the ongoing debate about high-mass star formation process. There are several proposed theories and among those, two are more favoured. They are thought to either form in the same way as the low-mass stars (core accretion) or through competitive accretion (\citealt{Bonnell2006}; \citealt{Myers2013}; \citealt{Tan2014}; \citealt{Motte2018}). Although both theories suggest existence of massive protostellar disks, the stability of such disks is debated (\citealt{Ahmadi2019}; \citealt{Johnston2020}). On the one hand, many works show that such massive disks fragment at a radius threshold of ${\sim}100-200$\,au (\citealt{Kratter2006}; \citealt{Krumholz2009}; \citealt{Oliva2020}). On the other hand, other studies show that disks with radii of 1000\,au can also form (\citealt{Kuiper2010}; \citealt{Kuiper2011}; \citealt{Klassen2016}; \citealt{Kuiper2018}). Interferometric observations show evidence for disks around massive young stellar objects (\citealt{Jimenez2012}; \citealt{Sanchez2013}; \citealt{Hirota2014} \citealt{Hunter2014}; \citealt{Johnston2015}; \citealt{Zapata2015}; \citealt{Ilee2016}; \citealt{Cesaroni2017}; \citealt{Maud2019}; \citealt{Bogelund2019}; \citealt{Williams2022}). Disk masses are found to be around $3-12$\,M$_{\odot}$, disk radii around $ 800-2500$\,au and protostellar masses around $ 20-70$\,M$_{\odot}$. Given these observations, disks around massive protostars seem to be common rather than an exception.

A final difference of high-mass protostars from their low-mass counterparts is that they may host an HII region. HII regions are divided in different categories depending on their extents. In this work the most relevant categories are the hypercompact (HC) and ultracompact (UC) HII regions. They are defined to have extents of $\lesssim 10300$\,au (\citealt{Kurtz2005}; \citealt{Hoare2007}) and $\lesssim 20600$\,au (\citealt{Wood1989}; \citealt{Hoare2007}), respectively. Therefore, the effect of such regions needs to be considered.


In this paper we address the question whether massive protostellar disks and optically thick continuum can explain the lack of methanol emission toward high-mass protostars. To answer this question we study an envelope-only and an envelope-plus-disk model following a similar method to \cite{Nazari2022}. We calculate the temperature and methanol emission by detailed radiative transfer modeling. For both models we consider optically thin and thick dust at mm wavelengths and parametrize the methanol abundance in the disk and the envelope. 

The main difference between the models in \cite{Nazari2022} and those in this work is the viscous heating that is included in the disk of high-mass protostars while this was not considered in \cite{Nazari2022}. Moreover, the region of the parameter space that this work considers includes higher envelope masses and protostellar luminosities to match those of observations of high-mass protostellar systems.

In Sect. \ref{sec:methods} we summarize our methods. Section \ref{sec:results} presents the results, in particular the effect of a disk on the temperature structure and the resulting methanol emission. Moreover, we explore the temperature inversion effect suggested by \cite{Barr2020}. We discuss our findings in Sect. \ref{sec:discussion}, especially our results are compared with observations and the effect of HII regions is discussed. Finally, we present our conclusions in Sect. \ref{sec:conclusions}.

\section{Methods}
\label{sec:methods}
\subsection{Physical structure and abundance}
\label{sec:phycical_structure}

In this work two models are considered: an envelope-only model and an envelope-plus-disk model. The two models have the same physical structure as in \cite{Nazari2022} and thus their details are only briefly stated here. The gas density structures of the two models are presented in Fig. 1. The envelope-only model has a power-law relation between gas density and radius (in spherical coordinates $r$) with its power fixed to -1.5 (i.e., $\rho_{\rm g}\propto r^{-1.5}$). This value is chosen to be consistent with observations of massive protostellar envelopes (\citealt{vanderTak2000}; \citealt{Gieser2021}). The envelope-plus-disk model consists of a flattened envelope density structure with an embedded disk. The flattened envelope model has a gas density structure following \cite{Ulrich1976}. The disk density follows a power-law in (cylindrical) $R$ and a Gaussian profile in $z$ direction (\citealt{Shakura1973}; \citealt{Pringle1981}). We assume a disk aspect ratio ($H/R$) of 0.2 similar to \citealt{Nazari2022} (also see \citealt{Harsono2015}). A gas-to-dust mass ratio of 100 is assumed for both models. An outflow cavity is carved for both models in the same way as done in \cite{Nazari2022}. The outflow cavity has a curved opening with total hydrogen nuclei number density fixed to 10$^3$\,cm$^{-3}$ where $\cos{\theta_{0}} > 0.95$. Here $\theta_{0}$ is the latitude of the particle at its initial location in the envelope. The curved opening angle is important for UV penetration into the envelope (\citealt{Bruderer2009}).

\begin{figure*}
    \centering
    \includegraphics[width=0.8\textwidth]{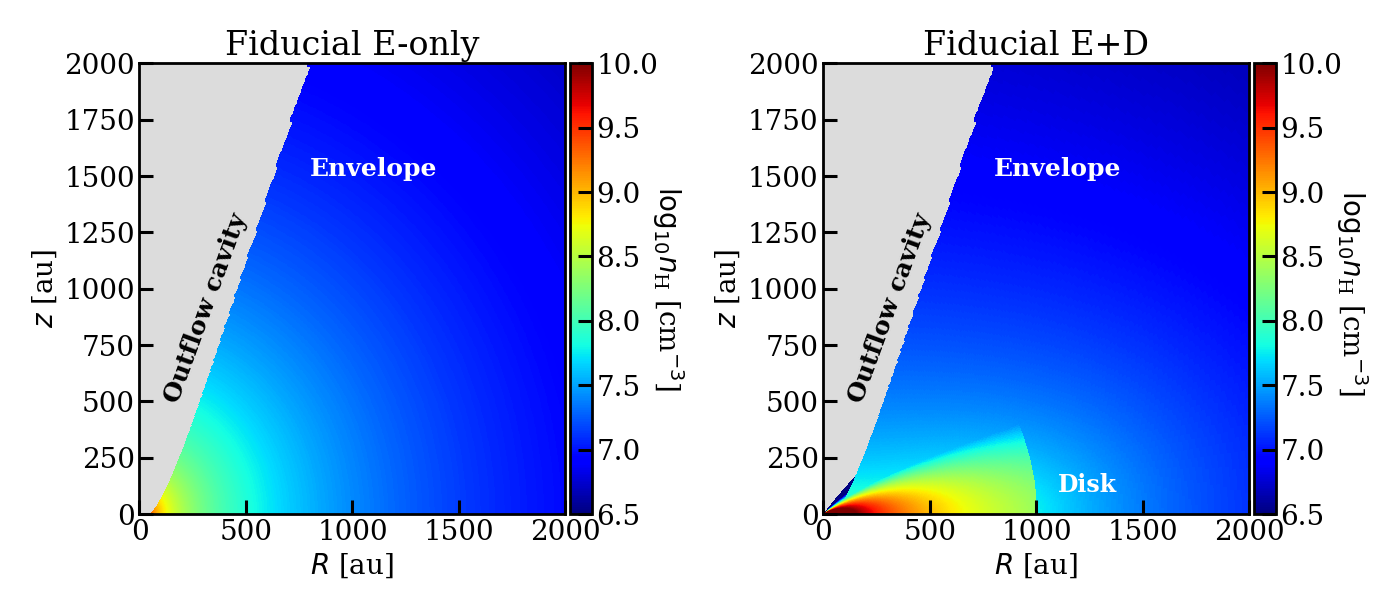}
    \caption{Gas density profiles. Two dimensional total hydrogen nuclei number density for the fiducial envelope-only model (left). The same but for the fiducial envelope-plus-disk model (right). The outflow cavities in this and subsequent figures are masked gray.} 
    \label{fig:H_density}
\end{figure*}

\begin{table*}[t]
  \centering
  \caption{Parameters of the models}
  \label{tab:params}
  \begin{tabular}{lllll}
  \toprule
  \toprule
  Parameter [unit] & Envelope-only  & Envelope-plus-disk & Description  \\
  \midrule
    $r_{\rm in}$ [au] & 10 & 10 & The inner radius\\
    $r_{\rm out}$ [au]  & $5\times10^{4}$ & $5\times10^{4}$ & The outer radius of the envelope\\
    $M_{\rm E}$ [M$_{\odot}$] & 50, 150, \textbf{300}, 550, & 50, 150, \textbf{300}, 550, & Envelope mass\\& 800, 1000 & 800, 1000 & \\
    $R_{\rm D}$ [au] & -- & 300, 500, \textbf{1000}, 1500, & Disk radius\\&&  2000, 2500 & \\
    $M_{\rm D}$ [M$_{\odot}$] & -- & 0.27,  0.75,  \textbf{3.}  ,  6.75, & Disk mass\\&& 12.  , 18.75\\
    $\dot{M}$ [M$_{\odot}$ yr$^{-1}$] & -- & $3.3 \times 10^{-4}$, $9.8 \times 10^{-4}$, \textbf{2.0}$\times$ \textbf{10}$^{-3}$, & Mass accretion rate\\&& $3.6 \times 10^{-3}$, $5.2 \times 10^{-3}$, $6.5 \times 10^{-3}$ &\\
    $T_{\star}$ [K]  & 40000 & 40000 & Protostellar temperature\\
    $M_{\star}$[M$_{\odot}$]  & 30 & 30 & Protostellar mass\\
    $L$ [L$_{\odot}$] & $5\times10^2$, 5$\times$10$^3$, \textbf{10$^4$}, 5$\times$10$^4$, & $5\times10^2$, 5$\times$10$^3$, \textbf{10$^4$}, 5$\times$10$^4$, & Bolometric luminosity\\& $5\times10^5$ , $5\times10^6$ & $5\times10^5$ , $5\times10^6$ & \\
    \bottomrule
  \end{tabular}
  \tablefoot{The parameters of the fiducial model are highlighted with bold face. The disk masses are varied such that $M_{\rm D}/R_{\rm D}^{2}$ (an approximation to the disk surface density) stays constant assuming a fiducial disk mass of 3\,M$_{\odot}$. The centrifugal radius is fixed to 500\,au (see Eq. 2 in \citealt{Nazari2022} for its effect).}
\end{table*}

Envelope masses of the modelled protostars are varied between 50\,M$_{\odot}$ and 1000\,M$_{\odot}$ following single-dish observations of the extended envelopes (\citealt{vanderTak2000}; \citealt{vanderTak2013}; \citealt{Benz2016}; \citealt{Konig2017}; \citealt{Pitts2022}). The bolometric luminosities are varied between $5\times 10^2$\,L$_{\odot}$ and $5 \times 10^6$\,L$_{\odot}$ (e.g. see \citealt{Lumsden2013}; \citealt{Elia2017} for the observed values for high-mass objects). We note that the luminosity range and envelope mass range assumed here includes the range that is often referred to as intermediate mass protostars ($L \lesssim 10^4$\,L$_{\odot}$ and $M_{\rm E} \lesssim 100$\,M$_{\odot}$). However, we keep these values for completeness. The disk radii for the envelope-plus-disk models span a range between 300\,au and 2500\,au following the disks observed around O- and B-type protostars (\citealt{Hunter2014}; \citealt{Johnston2015}; \citealt{Ilee2016}; \citealt{Ilee2018}; \citealt{Zhang2019}; \citealt{Sanna2019}; \citealt{AnezLopez2020}). The disk masses are varied such that $M_{\rm D}/R_{\rm D}^2$ stays constant. The disk mass for the fiducial model with disk radius of 1000\,au is assumed as 3\,M$_{\odot}$ (resulting in a disk mass range of ${\sim} 0.3-19$\,M$_{\odot}$). This value is chosen to be consistent with the observed massive disks around O- and B-type protostars (references given above). The central protostar's mass and temperature are fixed to 30\,M$_{\odot}$ and 40000\,K. In Sect. \ref{sec:caveats} we discuss the effect of changing the protostar's mass and temperature on methanol emission. The envelope's outer radius is fixed to $5 \times 10^4$\,au (\cite{vanderTak2000}; \citealt{Shirley2002}; \citealt{Pitts2022}). The inner radius is taken as 10\,au. However, because the temperature for some models (especially those with the highest luminosities) at radii between 10\,au and 20\,au goes above 2000\,K (upper limit on dust sublimation temperature), the methanol abundance is set to zero in the inner 20\,au. This assumption does not change the integrated methanol flux considered in the paper. All these parameters are summarised in Table \ref{tab:params}. We do not include an HII region in our models. However, its effect on methanol emission is discussed in Sect. \ref{sec:HII}. The fiducial envelope-only and envelope-plus-disk models throughout this work are defined to be those with $M_{\rm env}$ of $300$\,M$_{\odot}$, $L$ of $10^4$\,L$_{\odot}$, $R_{\rm D}$ of $1000$\,au, and $M_{\rm D}$ of $3$\,M$_{\odot}$ with small $\kappa_{\rm dust,\, mm}$ (see the highlighted values in Table \ref{tab:params}).

Methanol abundance in the disk and the envelope are calculated by balancing adsorption and thermal desorption (\citealt{Hasegawa1992}). The binding energy of methanol is assumed as 3820\,K (\citealt{Penteado2017}). Total methanol abundance ($X_{\rm gas}+X_{\rm ice}$) with respect to total hydrogen in the envelope is taken to be $10^{-6}$, with a minimum of $10^{-9}$ outside of the snow surface for $X_{\rm gas}$, following what \cite{Nazari2022} used for low-mass protostars (also see \citealt{Drozdovskaya2015}). This is justified given that the methanol ice abundances of low- and high-mass young stellar objects with respect to hydrogen are similar (\citealt{Oberg2011}; \citealt{Boogert2015}). In the disk, the total ice and gas abundance of methanol is assumed to be $10^{-8}$ with a minimum of $10^{-11}$ for $X_{\rm gas}$. These values are based on the modeling and observational works of low-mass protostars (\citealt{Walsh2014}; \citealt{Booth2021}) and mimic the potential effect from shocks that can destroy methanol. We note that the methanol abundance found by \cite{Bogelund2019} in the envelope/disk of AFGL 4176 is ${\sim}10^{-5}-10^{-6}$. This value could be overestimated because of continuum optical thickness. However, we explore higher assumed disk abundances in Sect. \ref{sec:caveats} and explain its effect on methanol emission. In our models chemical evolution of methanol in the disk and envelope are not included directly to focus on the effect of disk on methanol emission. These effects, however, are included implicitly by parametrizing the methanol abundance based on the previous observations and chemical models. Further effects from chemical evolution are discussed in Sect. \ref{sec:discussion}.

The photodissociation regions of methanol around the cavity walls are calculated in the same way as \cite{Nazari2022}. In short, we assume that methanol is photodissociated in the regions alongside the outflow cavity wall where $\tau_{\rm UV} < 3$ and hence its abundance is set to zero in these regions. The opening angle for the outflow cavity considered here is ${\sim} 20$ degrees narrower than that in \cite{Bruderer2009}. However, as discussed in \citet{Bruderer2009,Bruderer2010} the warm ($T > 100$\,K) mass only changes by less than a factor three for different cavity shapes and opening angles. Moreover, our photodissociation regions (where $\tau_{\rm UV} < 3$) for the low mm opacity dust grains have similar extents to those in \cite{Bruderer2009} (see Fig. \ref{fig:meth_abund_M50} and their Fig. 3).

\subsection{Temperature calculation}

We use RADMC-3D (\citealt{Dullemond2012}) version 2.0\footnote{\url{http://www.ita.uni-heidelberg.de/~dullemond/software/radmc-3d}} to calculate the dust temperature in the envelope and the disk. The same two dust distributions as \cite{Nazari2022} are considered (see their Appendix A for $\kappa_{\rm abs}$ as a function of wavelength). One with $\kappa_{1\,\rm mm}\simeq 0.2$\,cm$^2$\,g$^{-1}$ and another with $\kappa_{1\,\rm mm}\simeq 18$\,cm$^2$\,g$^{-1}$ to include the two extreme cases of low and high dust opacity at mm wavelengths, representing small and large grains respectively. The two dust distributions are referred to as low mm opacity and high mm opacity dust for the rest of this work. 

The grids for both envelope-only and envelope-plus-disk models are logarithmically spaced with 1000 and 400 grid points in the $r$ and $\theta$ direction, respectively. Moreover, $10^6$ photons are used for the temperature calculation. The number of grid cells and photons are chosen to produce accurate temperatures while maintaining a reasonable computation time. 

The models are exactly the same in this work and in \cite{Nazari2022} except for the viscous heating included in the disk for high-mass protostars here (also see \citealt{Harsono2015} for viscous heating included in low-mass protostellar disks) and the stellar spectrum assumed. Here these two differences are explained.

\begin{figure*}
    \centering
    \includegraphics[width=\textwidth]{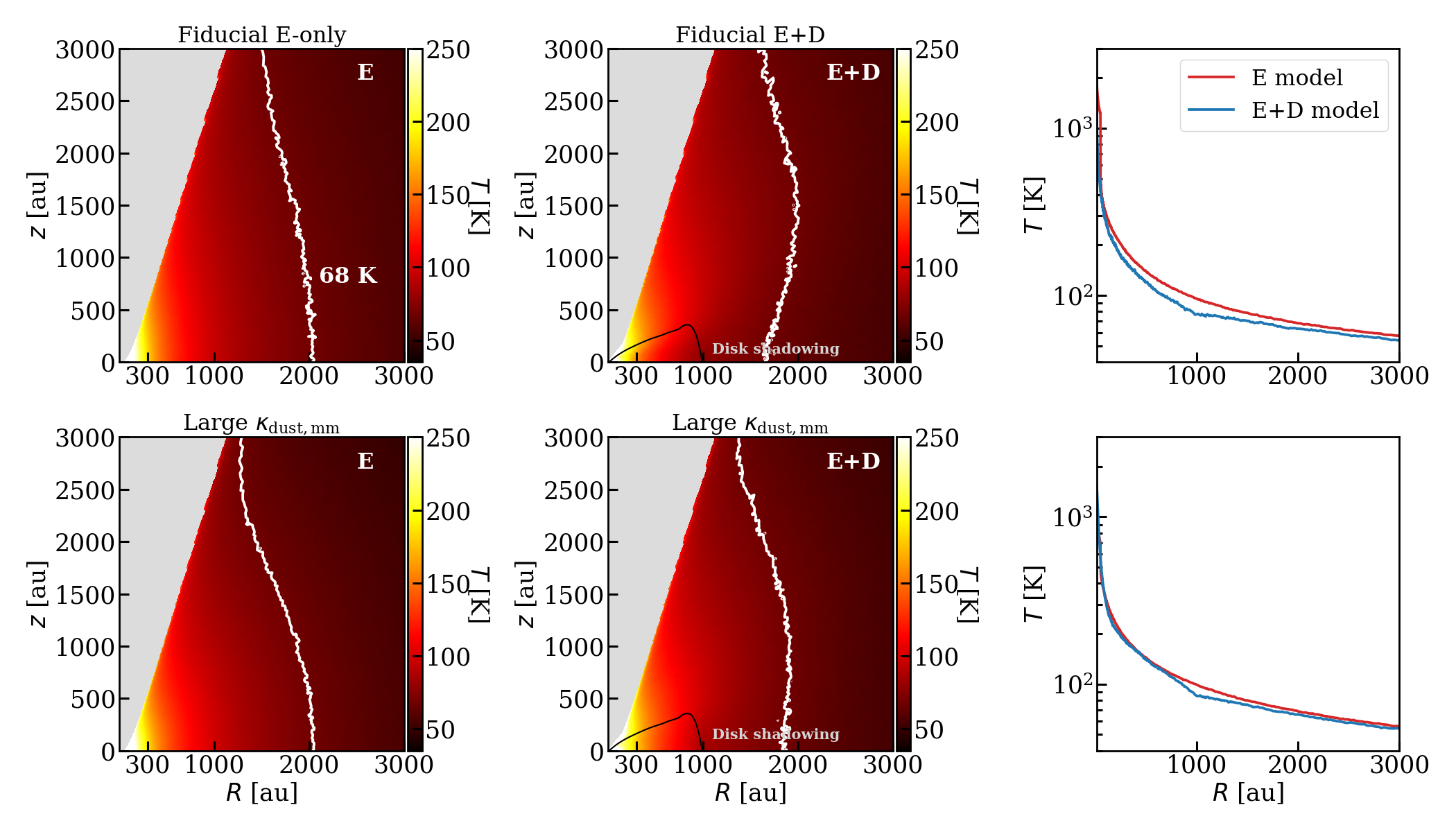}
    \caption{Temperature structure of the fiducial envelope-only and envelope-plus-disk. Top row shows the models with low mm opacity dust ($\kappa_{1\,\rm mm}\simeq 0.2$\,cm$^2$\,g$^{-1}$) and the bottom row shows those with high mm opacity dust ($\kappa_{1\,\rm mm}\simeq 18$\,cm$^2$\,g$^{-1}$). The right column shows a temperature cut for the various models at $z=0$\,au. The white contours show where temperature is 68\,K (roughly where methanol sublimates from the grains). The black contours show the approximate position of the disk.} 
    \label{fig:temp}
\end{figure*}

The reason to include viscous heating is that massive disks have high accretion rates (\citealt{Beuther2017}). Starting from the disk gas surface density steady state solution, we can write viscosity as (\citealt{Pringle1981}; \citealt{Lodato2008}) 


\begin{equation}
    \nu = \frac{\dot{M}}{3\pi \Sigma} \left(1-\sqrt{\frac{R_{\rm in}}{R}}\right).
    \label{eq:visc}
\end{equation}


\noindent where $\Sigma$ is the gas surface density, $\dot{M}$ is the accretion rate, $R$ is the radius in cylindrical coordinates and $R_{\rm in}$ is the inner radius of the disk. 

The viscous torques are important for angular momentum transfer throughout the disk to allow accretion but they also cause energy being dissipated. One can find the power lost per unit volume by the viscous torques in the disk using

\begin{equation}
    Q(R,z) = \frac{-G(R,z)\Omega^{\prime}}{2\pi R},
    \label{eq:Q_init}
\end{equation}

\noindent where $G(R,z)$ is the torque exerted by viscosity per unit length and $\Omega^\prime$ is $d\Omega/dR$ with $\Omega$ being the angular velocity. The torque per length is given by

\begin{equation}
    G(R,z) = -2\pi \nu \rho R^3 \Omega^\prime.
    \label{eq:torque}
\end{equation}




We can substitute $\nu$ from Eq. \eqref{eq:visc} into Eq. \eqref{eq:torque} and then substitute the resulting $G(R,z)$ into Eq. \eqref{eq:Q_init} to find the power dissipated by viscosity per unit volume as

\begin{equation}
    Q(R,z) = \frac{3}{4\pi} \frac{\dot{M} \rho \Omega^2}{\Sigma} (1-\sqrt{\frac{R_{\rm in}}{R}}).
    \label{eq:Q_final}
\end{equation}

\noindent This expression is found for each grid cell and is included as an extra heating source in the temperature calculation of RADMC-3D. Although the viscous torque (Eq. \ref{eq:torque}) depends on $\nu$, that itself depends on mass accretion rate. An assumption that is made here is that the disk is viscous enough to deliver all the accretion rate it receives from the envelope and there is no pile up of material at the envelope-disk intersect. Hence, Eq. \eqref{eq:Q_final} only depends on mass accretion rate. In other words, we do not vary $\nu$ directly but by varying $\dot{M}$ we take into account various viscous torques. Therefore, we only refer to mass accretion rate in our models which is calculated self-consistently for the density profile and the parameters considered ($\dot{M}\,[\rm{M_{\odot} yr^{-1}}]/(2 \times 10^{-3}) \simeq M_{\rm env}\,[\rm{M_{\odot}}]/300$). The accretion rates are between $3.3\times 10^{-4}$\,M$_{\odot}$\,yr$^{-1}$ and $6.5\times 10^{-3}$\,M$_{\odot}$\,yr$^{-1}$ in this work.

The stellar spectra with the luminosities given in Table \ref{tab:params} are not simple blackbodies as assumed in \cite{Nazari2022}. This is because the central massive protostar has an effective temperature of 40000\,K and hence ${\sim} 40\%$ of the photons in the stellar blackbody spectrum will ionise hydrogen. Because in reality these photons will get absorbed by the hydrogen atoms before any direct contact with dust and later are re-radiated at longer wavelengths, we alter the blackbody spectrum to simulate this effect. We assume that these photons are later re-emitted at the Lyman-$\alpha$ wavelength with a width of 18\,km\,s$^{-1}$ for a Gaussian profile ($\rm{FWHM} = 2 \sqrt{2\ln{2}} \times 18$). This is approximately equal to the line width found from thermal broadening at temperature of 40000\,K. We discuss the effect of completely removing the ionising photons from the stellar blackbody as a case producing a lower limit on methanol emission in Sect. \ref{sec:caveats}.


\subsection{Line emission calculation}

The line emission is calculated using RADMC-3D version 2.0. The molecular data are taken from the Leiden Atomic and Molecular Database (downloaded on 16$^{\rm th}$ of February 2022; \citealt{Schoier2005}; \citealt{vanderTak2020}). The line properties such as frequency, upper energy level and Einstein A coefficient are taken from the Cologne Database for Molecular Spectroscopy (CDMS; \citealt{Muller2001}; \citealt{Muller2005}). We calculate the emission from one of the strong methanol lines covered in the ALMA Evolutionary study of High Mass Protocluster Formation in the Galaxy (ALMAGAL) survey (\citealt{vanGelder2022}). This is to compare our results with those observations. The chosen methanol transition has J K L M - J K L M quantum numbers equal to 4 2 3 1-3 1 2 1 and has a frequency of 218.4401\,GHz ($E_{\rm up}=45.5$\,K, $A_{\rm i,j} = 4.7 \times 10^{-5}$\,s$^{-1}$). Because this line has a similar upper state energy and Einstein $A$ coefficient to the line used in \cite{Nazari2022}, where the local thermodynamic equilibrium (LTE) assumption was found to be valid, we assume LTE conditions. This is well justified since the densities we consider here are even higher than in the low-mass case.


The ray tracing is done in the same way as \cite{Nazari2022} with a spectral resolution of 0.2\,km\,s$^{-1}$. The source is assumed to be located at a distance of 4\,kpc (typical distance of high-mass protostars; e.g. \citealt{Mege2021}). Gas and dust are included in ray tracing and subsequently the lines are continuum subtracted before calculating the integrated line fluxes. The emission is integrated over a $2\arcsec$ area. This corresponds to a source diameter of 8000\,au for a source located at 4\,kpc. The $2\arcsec$ is chosen to simulate the angular resolution of surveys of massive protostars such as ALMAGAL and is large enough to include the disk and the hot core region where methanol is sublimated for most models. The models with the highest luminosities considered often are hot enough to sublimate methanol up to the outer radii assumed here.

In the envelope we assume a turbulent velocity of 2\,km\,s$^{-1}$ (slightly larger than what was assumed in \citealt{Nazari2022} for low-mass protostars as 1\,km\,s$^{-1}$). This turbulent velocity produces a line emission with full width at half maximum (FWHM) of ${\sim} 4$\,km\,s$^{-1}$. The FWHM of lines toward high-mass protostars are on average larger than their low-mass counterparts (e.g., \citealt{Nazari2021}; \citealt{Nazari2022ALMAGAL}). In the disk a turbulent velocity of 0.1\,km\,s$^{-1}$ and Keplerian velocity are assumed. Because double-peaked line profiles for methanol are not regularly observed, no free-fall velocity is assumed in the envelope. As discussed in \cite{Nazari2022} the inclusion of free-fall velocity should not change the main conclusions when focusing on integrated line fluxes.


\section{Results}
\label{sec:results}
In this section the main results are explained. Most importantly, we discuss the temperature structure and the resulting methanol emission.

\subsection{Temperature}
\label{sec:temp}
\subsubsection{General structure}
\label{sec:gen_structure}



\begin{figure*}
    \centering
    \includegraphics[width=16cm]{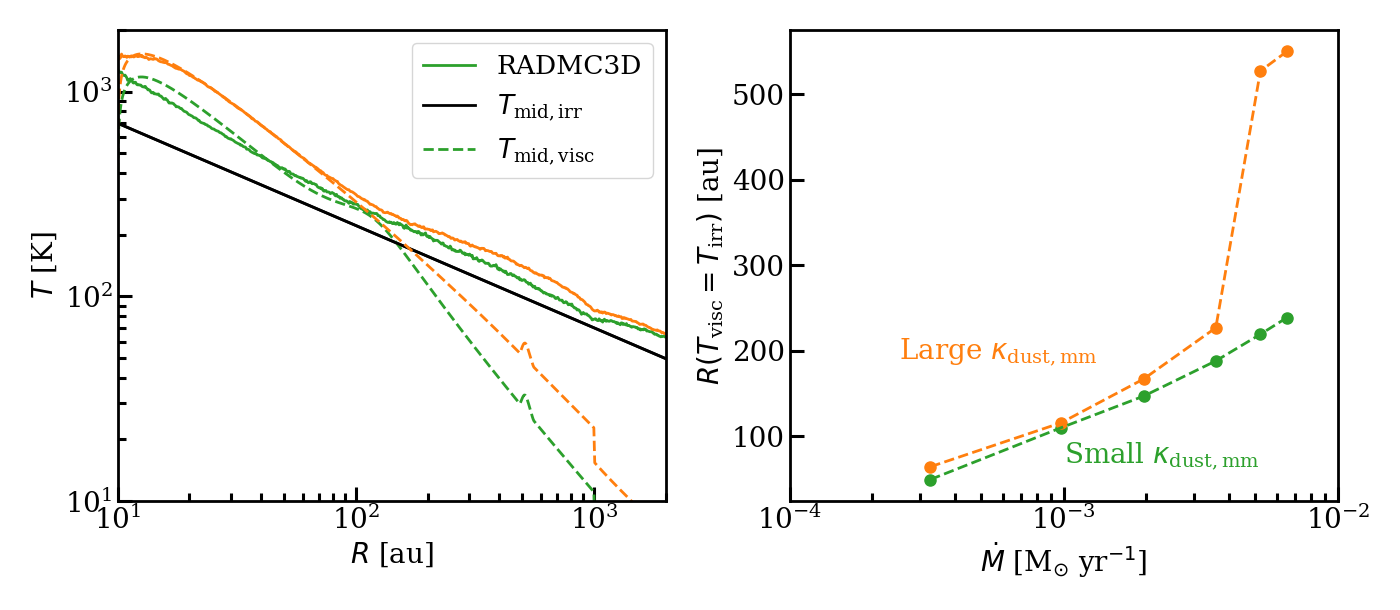}
    \caption{Left: Comparison of the mid-plane temperature calculated by RADMC-3D for the fiducial envelope-plus-disk model (orange and green solid lines), the same calculated from viscous heating analytically (dashed lines) and found from passive heating from the protostar analytically (black solid line). Green shows dust with small $\kappa_{mm}$ and orange shows that for large $\kappa_{mm}$. Right: The radius at which the analytical mid-plane temperature from viscous heating equals that from passive heating plotted against mass accretion rate for the fiducial model and that with large mm opacity dust.} 
    \label{fig:temp_compare}
\end{figure*}

Figure \ref{fig:temp} shows the temperature structure of the fiducial ($M_{\rm env} = 300$\,M$_{\odot}$, $L=10^4$\,L$_{\odot}$, $R_{\rm D} = 1000$\,au, and $M_{\rm D} = 3$\,M$_{\odot}$ with small $\kappa_{\rm dust,\, mm}$) envelope-only and envelope-plus-disk models along with those with large $\kappa_{\rm dust,\, mm}$. The temperatures found in the disk (${\sim} 150-200$\,K) agree with what has been observed and assumed previously for disks of massive protostars (\citealt{Johnston2015}; \citealt{Izquierdo2018}; \citealt{Maud2019}). The effect of shadowing behind the disk is observed when the dust has low and high mm opacity. However, this phenomenon does not have as strong an effect on the temperature structure as for low-mass protostars (\citealt{Nazari2022}) for the envelope mass and luminosity of the fiducial model. Moreover, the disk mid-plane is hot due to viscous heating which is in contrast to the case of low-mass protostars (\citealt{Nazari2022}). It is important to note that viscous heating is only effective in changing the temperature structure in the most inner radii (${\lesssim} 100$\,au; also see Sect. \ref{sec:inversion}). These result in similar temperature structures between models with and without a disk while the models with a disk have slightly lower temperatures due to disk shadowing (see where the white contours cross the x-axis in Fig. \ref{fig:temp}).

There is little temperature difference between the low mm opacity dust (top row) and high mm opacity dust (bottom row) of Fig. \ref{fig:temp}. In fact toward $z=0$\,au (mid-plane) the envelope-plus-disk model has higher temperatures when the dust has large $\kappa_{\rm mm}$ (compare the white contours in the middle column). This is surprising at first because from the findings of \cite{Nazari2022} it is expected that the high mm opacity dust absorbs UV and optical light poorly and hence is colder than low mm opacity dust. Moreover, once they absorb the UV and optical photons they re-emit more efficiently at longer wavelengths again making the region colder. However, this is not what is seen here which is more apparent in the disk mid-plane.

In the envelope-plus-disk models, this is because of the balance between viscous heating increasing the temperature in the model with high mm opacity dust, and the effects mentioned above (low $\kappa_{\rm UV}$ plus high $\kappa_{\rm mm}$) lowering the temperature in the same model. Once viscous heating is included in the disk, the temperature of optically thick regions of the disk (i.e., the dense mid-plane) depend on the dust optical depth (\citealt{DAlessio1998}). This is the reason that after including viscous heating the disk mid-plane has a hotter temperature than its surface by a factor ${\sim} \left(\frac{3}{4}\tau\right)^{1/4}$, where $\tau$ is the dust optical depth which is proportional to the Rosseland mean opacity over the wavelengths longer than ${\sim}0.1$\,$\mu$m (see Appendix \ref{sec:pass_visc}; \citealt{DAlessio1998}; \citealt{Armitage2010}). Therefore, for the dust distribution with high mm opacity (which also has a higher Rosseland mean opacity) the mid-plane temperature should be higher than the low mm opacity dust as seen from the white contours in the middle panel of Fig. \ref{fig:temp}.



\subsubsection{Heating sources}
\label{sec:heating_sources}
There are two sources of heating in our envelope-plus-disk models: radiation from the star (passive heating) and heating due to viscosity. In this section we quantify the effect from the two heating sources and compare the analytical solutions with the results from RADMC-3D models. Appendix \ref{sec:pass_visc} presents the formulae to calculate viscous heating and passive heating in the disk.

The left panel of Fig. \ref{fig:temp_compare} presents a comparison between the results from RADMC-3D, the analytical disk mid-plane temperature profile resulted from viscous heating ($T_{\rm mid,\,visc}$; Eq. \ref{eq:T_mid_visc} ), and that from passive heating ($T_{\rm mid, irr}$; Eq. \ref{eq:T_irr}). RADMC-3D results and the analytical solutions match well. The temperature profile in the inner disk is explained by viscous heating and that in the outer regions by passive heating. In particular, there is a radius threshold at which $T_{\rm mid, visc}$ (dashed lines) crosses $T_{\rm mid, irr}$ (black solid line). This radius is ${\sim} 200$\,au for the models shown in left panel of Fig. \ref{fig:temp_compare}. Inside this radius the contribution from viscous heating is larger than passive heating in the disk mid-plane. Quantitatively, for the large mm opacity dust there is a factor of about two difference between temperature resulting from viscous heating and that from passive heating at radii of around 10\,au.


The right panel of Fig. \ref{fig:temp_compare} presents the relationship between mass accretion rate and the threshold radius described above calculated from the analytical formulae given in Appendix \ref{sec:pass_visc}. This threshold can be found by equating $T_{\rm mid,\, irr}$ in Eq. \eqref{eq:T_irr} and $T_{\rm mid,\, visc}$ from Eq. \eqref{eq:T_mid_visc} . Figure \ref{fig:temp_compare} shows that increasing the mass accretion rate will increase the radius inside of which viscous heating is dominant. In other words, for lower mass envelopes or disks around lower mass stars (i.e., lower accretion rates) viscous heating is only effective in the inner regions of the disk (${\lesssim} 100$\,au) as expected from Eq. \eqref{eq:Q_final} (\citealt{DAlessio1998}; \citealt{Harsono2015}). More quantitatively, there is a factor ${\gtrsim} 2$ difference in threshold radius between the models with mass accretion rate of ${\sim}3.6 \times 10^{-3}$\,M$_{\odot}$\,yr$^{-1}$ and ${\sim}3 \times 10^{-4}$\,M$_{\odot}$\,yr$^{-1}$.

\begin{figure*}
    \centering
    \includegraphics[width=0.9\textwidth]{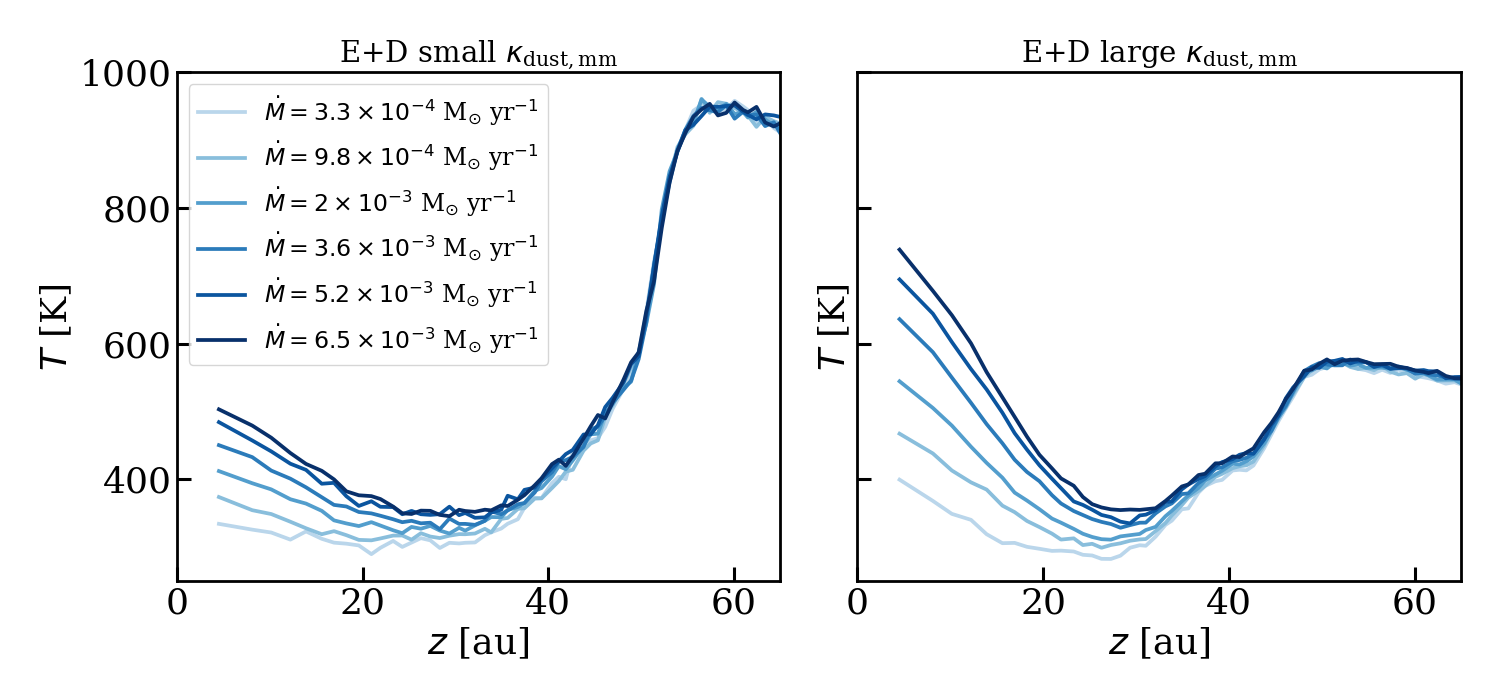}
    \caption{Vertical temperature cut at a radius of ${\sim}50$\,au. Left: The fiducial envelope-plus-disk model with varying envelope masses and thus various mass accretion rates. Right: The same as left but for the fiducial models with large $\kappa_{\rm dust,\, mm}$. The $z$ of ${\sim}50-60$\,au marks where roughly the outflow cavity wall is reached and hence the highest $z$ at the disk or envelope surface at radius of ${\sim}50$\,au.} 
    \label{fig:temp_vertical}
\end{figure*}

\begin{figure}
  \resizebox{\hsize}{!}{\includegraphics{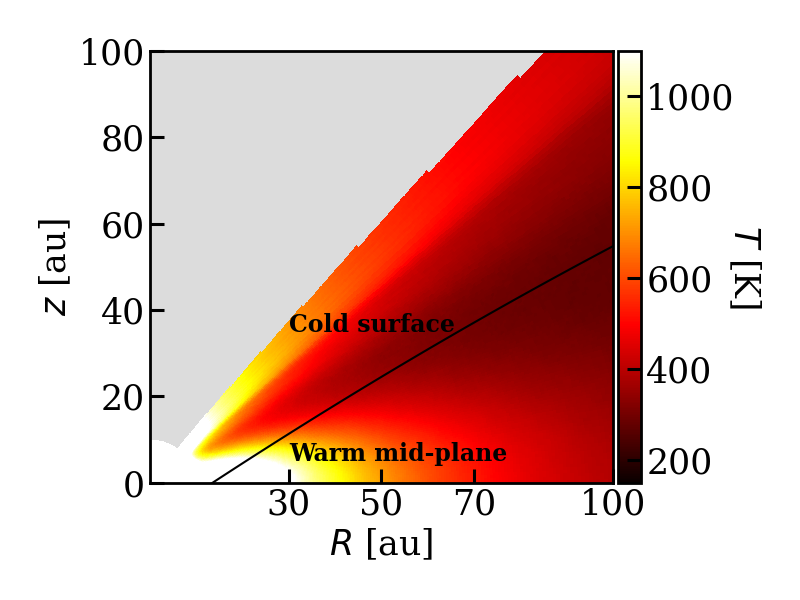}}
  \caption{Two dimensional map of temperature for the fiducial envelope-plus-disk model but with the difference that this model has large mm opacity dust and envelope mass of 800\,M$_{\odot}$ ($\dot{M} = 5.2 \times 10^{-3}$\,M$_{\odot}$\,yr$^{-1}$). The black contour shows the approximate location of the disk. In the inner disk ($R \lesssim 50-60$\,au) the mid-plane temperature is larger than the disk surface and envelope temperature.}
  \label{fig:temp_inver_2D}
\end{figure}

\begin{figure}
  \resizebox{\hsize}{!}{\includegraphics{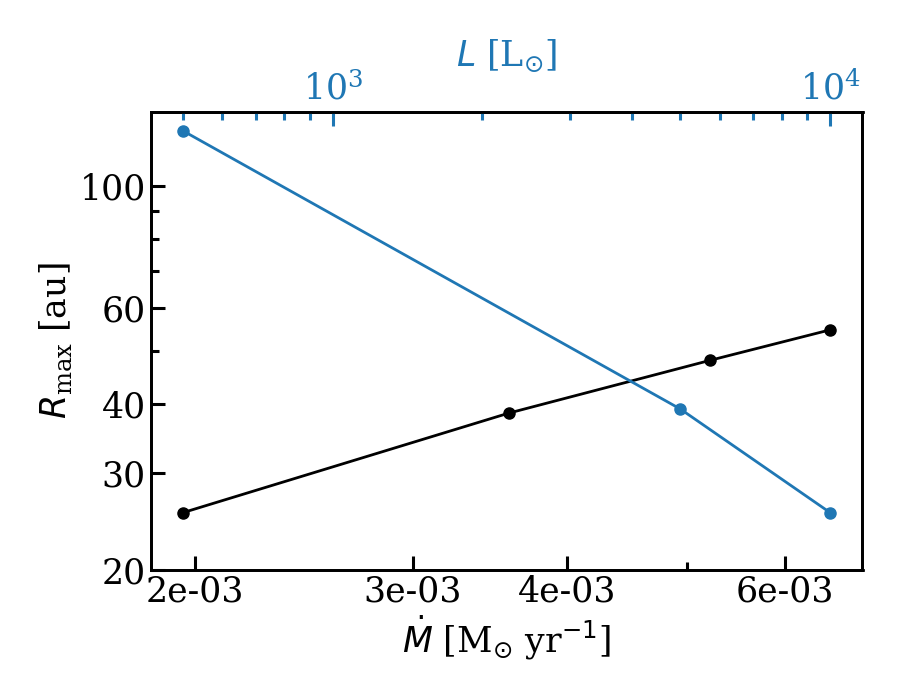}}
  \caption{The maximum radius at which the mid-plane temperature is larger than the surface temperature as a function of mass accretion rate (black) and luminosity (blue). The models shown here are fiducial envelope-plus-disk models with large grains (i.e., high mm opacity dust) where $M_{\rm E}$ ($\dot{M}$) changes for the black line and $L$ changes for the blue line.}
  \label{fig:T_surf_mid_analytic}
\end{figure}

\subsubsection{Vertical temperature inversion}
\label{sec:inversion}

High-mass protostellar disks have high accretion rates resulting in viscous heating in the disk. In particular, Sect. \ref{sec:heating_sources} explains and Fig. \ref{fig:temp_compare} shows where in the disk mid-plane, viscous heating is a more dominant source of heating than passive heating from the protostar. However, it is not clear whether viscous heating will cause higher temperatures in the disk mid-plane than the disk surface. Observations of mid-infrared absorption lines of CO, CS, HCN, C$_2$H$_2$, and NH$_3$ toward the inner radii of the potential disks around AFGL 2136 and AFGL 2591 suggest that the disk surface is colder than the disk mid-plane (\citealt{Barr2020}). In this section we explore this idea and investigate whether such temperature inversion is observed in our models.

Figure \ref{fig:temp_vertical} presents vertical cuts for the fiducial envelope-plus-disk model with varying envelope mass (i.e. accretion rates) and the same with large $\kappa_{\rm dust,\, mm}$. These cuts are made at ${\sim}50$\,au which is the radius at around which \cite{Barr2020} find the absorption lines originate. In Fig. \ref{fig:temp_vertical} the outflow cavity wall starts at $z\simeq50-60$\,au, indicating where the top surface of the disk or envelope is. So with these models a larger temperature at $z=0$\,au than the temperature just before hitting the cavity wall (i.e., $z\simeq50-60$\,au) is needed for vertical temperature inversion. The left panel of Fig. \ref{fig:temp_vertical} shows that for none of the accretion rates considered here such temperature inversion is observed when the dust has a low mm opacity (i.e., small dust). In other words, the temperature in the mid-plane is always smaller than the disk surface when the grains have low $\kappa_{\rm dust, \, mm}$. We note that the slight decrease in temperature seen (between $z=0$\,au and $z\simeq30$\,au) in left panel of Fig. \ref{fig:temp_vertical} is not enough for observations of absorption lines because still the temperature at $z=0$\,au is smaller than that at $z\simeq60$\,au.

\begin{figure*}
    \centering
    \includegraphics[width=0.8\textwidth]{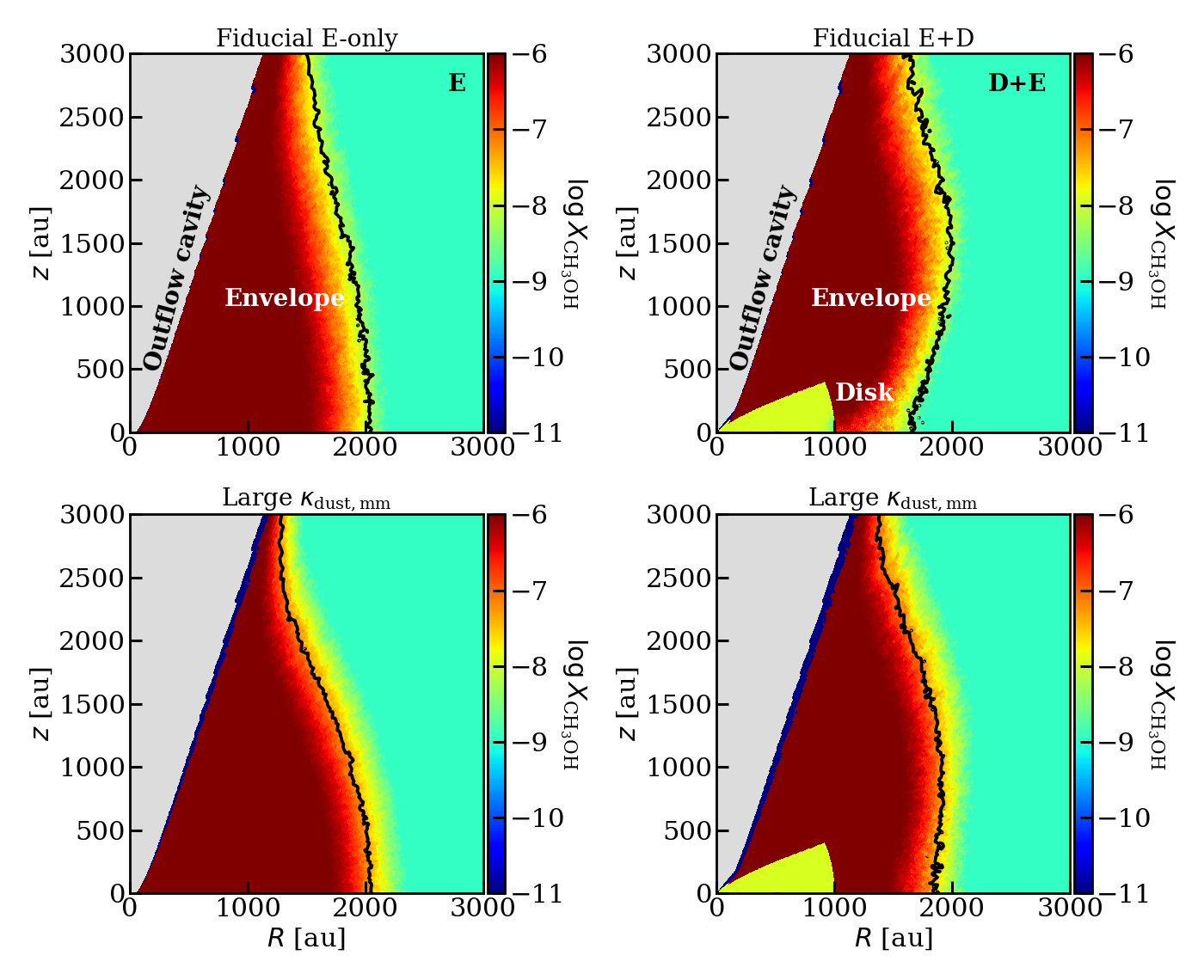}
    \caption{Gas-phase methanol abundance map for the fiducial envelope-only and envelope-plus disk models (top row) and those with large mm opacity dust (bottom row). The black contours show the 68\,K lines where methanol starts to be sublimated from the grains at the densities of these models.} 
    \label{fig:meth_abund}
\end{figure*}

The right panel of Fig. \ref{fig:temp_vertical} shows that for large dust distribution with high mm opacity the temperature inversion happens when the mass accretion rate is at least ${\sim} 3.6\times 10^{-3}$\,M$_{\odot}$\,yr$^{-1}$. Therefore,  a source with $R_{\rm D} = 1000$\,au, $L = 10^{4}$\,L$_{\odot}$ and high mm opacity dust distribution needs at least an envelope mass of ${\sim}550$\,M$_{\odot}$ ($M_{\rm env}\,[\rm{M}_{\odot}] \simeq 300/(2 \times 10^{-3})\, \dot{M}\,[\rm{M_{\odot} yr^{-1}}]$) to show larger temperatures in the disk mid-plane than the disk surface at a disk radius of 50\,au. Based on these results the models with high mm opacity dust reproduce the vertical temperature inversion suggested by the observations of \cite{Barr2020} better and hence might be more realistic. For sources with accretion rates $\gtrsim 5.2 \times 10^{-3}$\,M$_{\odot}$\,yr$^{-1}$ the difference between the temperature in the mid-plane and the disk surface is at least ${\sim} 100$\,K. This can differ when luminosities different from $L = 10^{4}$\,L$_{\odot}$ are considered or when the temperature cut is made at a different radius than 50\,au.

Figure \ref{fig:temp_inver_2D} shows a two dimensional temperature map of the fiducial envelope-plus-disk model but with large mm opacity dust and mass accretion rate of $5.2 \times 10^{-3}$\,M$_{\odot}$\,yr$^{-1}$. We can see more clearly in this figure that at radii below ${\sim}50-60$\,au the disk mid-plane temperature is hotter than that in the disk surface and envelope.

One can quantify this phenomenon further by comparing the analytical relations of mid-plane and surface temperatures of a disk. In Sect. \ref{sec:heating_sources} we discussed the analytical formulae for the mid-plane temperature (Eq. \ref{eq:T_mid}). The analytical relation for the disk surface temperature from viscous heating (Eq. \ref{eq:T_visc}) and passive heating (Eq. \ref{eq:T_surf_irr}) are also given in Appendix \ref{sec:pass_visc}. Therefore, using the total temperature in the mid-plane (Eq. \ref{eq:T_mid}) and that in the disk surface (Eq. \ref{eq:T_surf}) one can find the maximum radius ($R_{\rm max}$) at which temperature in the mid-plane is larger than that in the disk surface. This radius is dependent on the values of Rosseland mean opacity ($\kappa_{\rm R}$) and Planck mean opacity ($\kappa_{\rm P}$), which change as a function of radius. Therefore, an exact relation for R$_{\rm max}$ cannot be found and it needs to be solved numerically. However, assuming typical values of ${\sim}60$ and ${\sim}0.2$ (when dust grains have high mm opacity) for $\tau$ and $\epsilon$ which depend on $\kappa_{\rm R}$ and $\kappa_{\rm P}$ (see Appendix \ref{sec:pass_visc} for the complete definition) between radii of ${\sim}50$\,au and ${\sim}100$\,au, a simple approximate relation for $R_{\rm max}$ can be found

\begin{equation}
    R_{\rm max} \simeq \frac{3\dot{M}GM_{\star} \epsilon}{L (1-\epsilon \varphi)} (\frac{3}{4}\tau -1).
\end{equation}

\noindent Here, $\varphi$ is the flaring angle. This relation only gives a very rough estimate of $R_{\rm max}$ because $\tau$ for the fiducial model but with large dust grains varies between ${\sim}45$ and ${\sim} 90$ for radii between ${\sim}50$\,au and ${\sim}100$\,au. Moreover, these values would be different for the models with various $\dot{M}$ and $L$. 

\begin{figure*}
    \centering
    \includegraphics[width=0.8\textwidth]{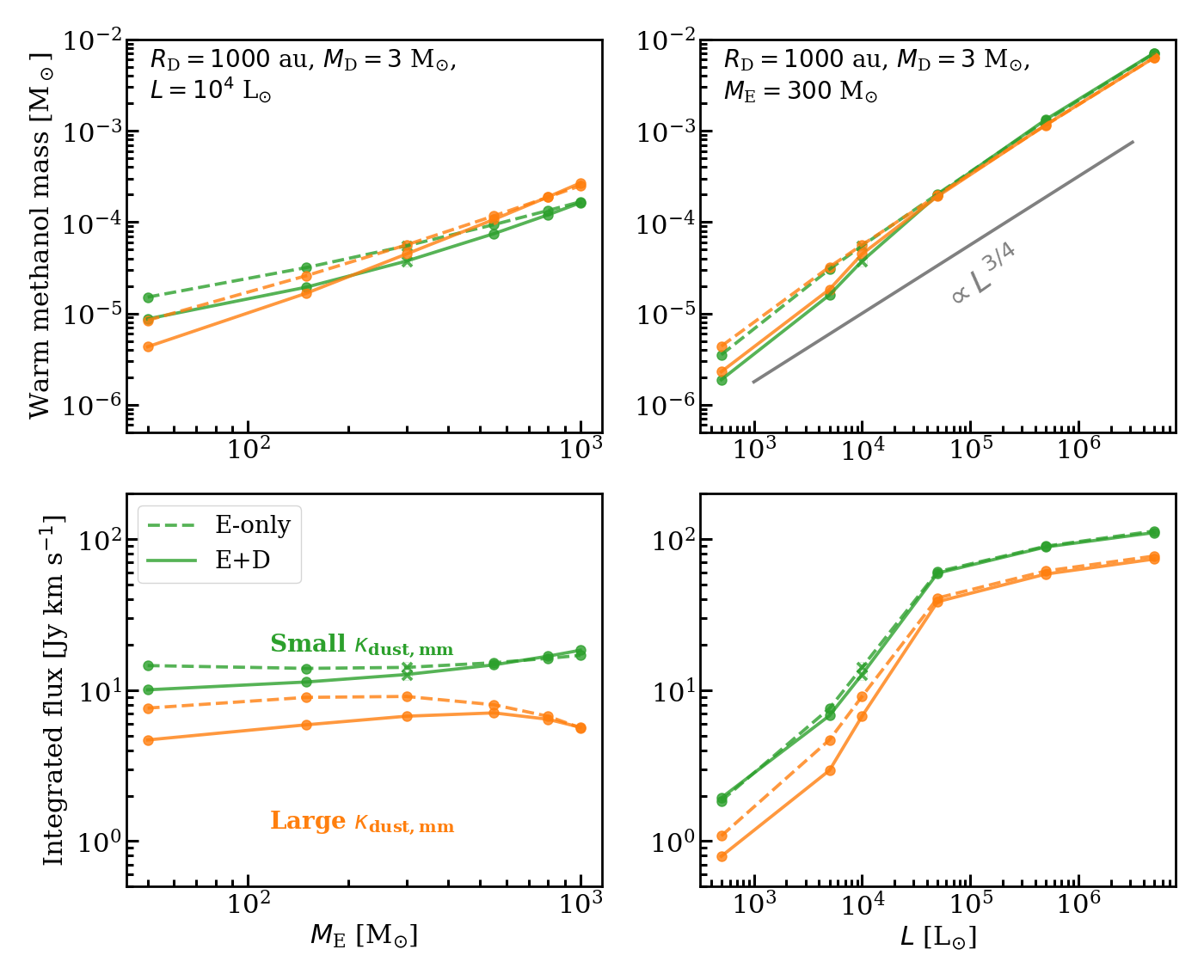}
    \caption{Warm methanol mass (top row) and integrated line fluxes (bottom row) for various models. The left column presents the models with varying envelope masses but constant luminosity of 10$^4$\,L$_{\odot}$ (i.e. varying accretion rates for the envelope-plus-disk models). The right column shows the models with varying bolometric luminosity but constant envelope mass of 300\,M$_{\odot}$ (i.e. constant accretion rate of $2\times 10^{-3}$\,M$_{\odot}$\,yr$^{-1}$). The parameters fixed for each column are printed on the top row plots. For example, where envelope mass is varied the disk radius is fixed to 1000\,au, disk mass is fixed to 3\,M$_{\odot}$ and luminosity is fixed to $10^4$\,L$_{\odot}$. Orange and green show the models with high and low mm opacity dust. The fiducial models are indicated by a cross. Solid and dashed lines present the envelope-plus-disk and envelope-only models, respectively. The gray solid line in the top right panel shows the analytical relation of warm methanol mass and luminosity which goes as $\propto L^{3/4}$ (\citealt{vanGelder2022}). Here this relation is normalized by an arbitrary value, hence, only its slope should be compared with the models. The integrated line fluxes are calculated after the lines are continuum subtracted and a source distance of 4\,kpc is assumed.} 
    \label{fig:mass_int}
\end{figure*}

Figure \ref{fig:T_surf_mid_analytic} presents $R_{\rm max}$ as a function of mass accretion rate (bottom axis in black) and luminosity (top axis in blue) for models with large grains (high mm opacity dust). In this figure $R_{\rm max}$ is calculated numerically with values for the mean opacities found iteratively as explained in Appendix \ref{sec:pass_visc}. At accretion rates below ${\sim} 2 \times 10^{-3}$\,M$_{\odot}$\,yr$^{-1}$ and luminosities above ${\sim}10^{4}$\,L$_{\odot}$ there is no radius (larger than the inner radius used in the models, i.e, 10\,au) at which mid-plane temperature is larger than the surface temperature. The same holds for all the models with small mm opacity dust grains. The reason that in this figure temperature inversion only occurs for the models with high mm opacity dust (also seen in Fig. \ref{fig:temp_vertical}) is that disk mid-plane temperature is proportional to the dust optical depth and thus Rosseland mean opacity (see Appendix \ref{sec:pass_visc}). Therefore, the temperature in the disk mid-plane is higher for the high mm opacity dust (which has a higher Rosseland mean opacity) than that for the low mm opacity dust.

In Fig. \ref{fig:T_surf_mid_analytic}, the maximum radius at which the temperature inversion occurs increases with increasing mass accretion rate. This is because viscous heating is proportional to $\dot{M}$ (see Eq. \ref{eq:Q_final}). Moreover, this maximum radius decreases as luminosity increases. This is because the increase in the disk surface temperature by passive heating (Eq. \ref{eq:T_surf_irr}) is steeper than the increase in the disk mid-plane temperature by passive heating (Eq. \ref{eq:T_irr}) as luminosity increases.

Focusing on the black line in Fig. \ref{fig:T_surf_mid_analytic} one can see that an accretion rate of at least ${\sim}2 \times 10^{-3}$\,M$_{\odot}$\,yr$^{-1}$ is needed for the temperature inversion. At accretion rates below this value there is no radius (above 10\,au) at which the mid-plane temperature is higher than the surface temperature. Moreover, this inversion only occurs up to radii of ${\sim} 30$\,au in the disk. In addition, the results from this figure are in-line with those from right panel of Fig. \ref{fig:temp_vertical}. For example Fig. \ref{fig:T_surf_mid_analytic} implies that for the temperature inversion to occur at radii of ${\sim}40-50$\,au an accretion rate of at least ${\sim} 3.6 \times 10^{-3}$\,M$_{\odot}$\,yr$^{-1}$ is needed which is the same as what is found from right panel of Fig. \ref{fig:temp_vertical}. Moreover, for luminosities above ${\sim} 10^{4}$\,L$_{\odot}$ no radius (above 10\,au) is found at which the temperature inversion happens which is also seen in our models (see Fig. \ref{fig:temp_vertical_L5e5}).

A caveat in this analysis is the decoupling of gas and dust temperature that is not considered here. A larger gas temperature than dust temperature in the disk surface is expected due to e.g., extra heating of the gas related to photoprocesses (\citealt{Kamp2004}; \citealt{Jonkheid2004}; \citealt{Bruderer2012}). Therefore, in reality the gas temperature in the mid-plane needs to be even higher than presented here for vertical temperature inversion to occur. This implies that high mm opacity dust models are more relevant and closer to reality than those with low mm opacity dust. To conclude, Figures \ref{fig:temp_vertical} and \ref{fig:T_surf_mid_analytic} show that the temperature inversion occurs in many of our models, especially those with large grains (i.e., high mm opacity dust). Therefore, a large area of the parameter space explored here agree with the conclusions of \cite{Barr2020}.

\begin{figure}
  \resizebox{\hsize}{!}{\includegraphics{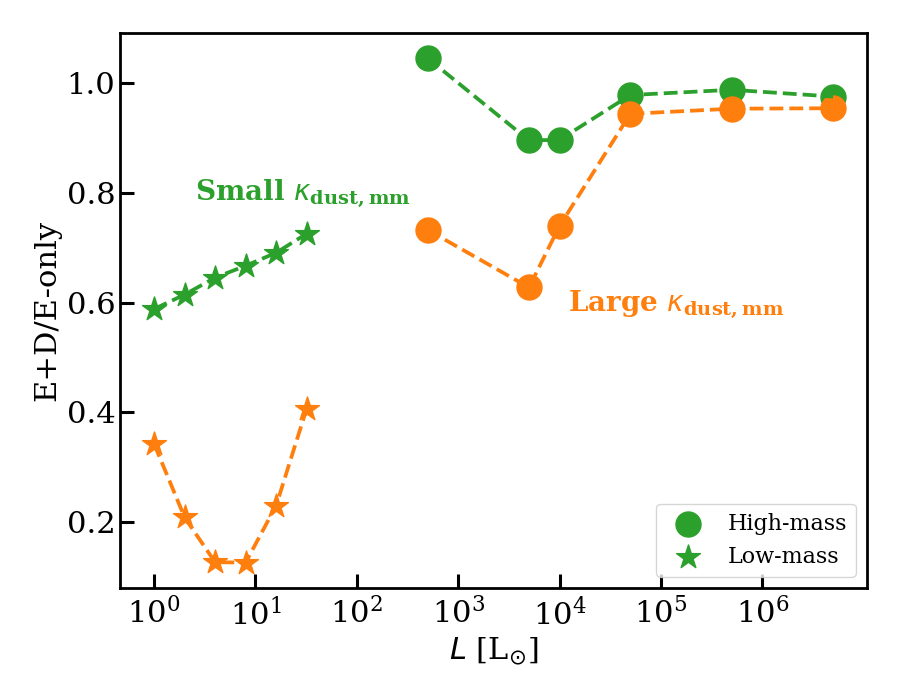}}
  \centering
  \caption{Comparison of methanol integrated fluxes between low- and high-mass protostars. The values for low-mass protostars are taken from \cite{Nazari2022}. The circles show the ratio of methanol integrated fluxes of fiducial envelope-plus-disk models with varying luminosity to those of envelope-only models, for high-mass protostars. The stars show the same for low-mass protostars. Green shows low mm opacity dust and orange high mm opacity dust. Low-mass disks are more effective in decreasing the methanol emission than high-mass ones.}
  \label{fig:low_high_comp}
\end{figure}

\subsection{Warm methanol mass and its emission}
\label{sec:warm_mass}

Figure \ref{fig:meth_abund} presents a methanol abundance map for our fiducial models and those with high mm opacity dust. Methanol is sublimated from the grains in all of the disk in our fiducial envelope-plus-disk model and that with large mm opacity dust grains. 

Moreover, the photodissociation regions next to the outflow cavity walls do not exist for the fiducial models and they are very thin in the fiducial models with high mm opacity dust due to high envelope densities. The photodissociation regions in low-mass protostars of \cite{Nazari2022} had an important effect in lowering the methanol mass and hence its emission toward low-mass protostars. However, smaller photodissociation regions were seen in low-mass protostars with envelope masses $\gtrsim 3$\,M$_{\odot}$ in \cite{Nazari2022} (see their Fig. E.3) due to the higher densities. Therefore, for high-mass protostars with larger envelope masses and densities than those in low-mass protostars, the photodissociation regions are expected to be smaller.

For completeness, Fig. \ref{fig:meth_emission} presents the resulting methanol emission and continuum subtracted line fluxes for the fiducial models. Due to the larger turbulent velocity the FWHM of the lines are larger (${\sim}4$\,km\,s$^{-1}$) than the low-mass protostars in \cite{Nazari2022} (${\sim} 2$\,km\,s$^{-1}$). The line emission has a higher peak (by a factor ${\sim}1.6$) when the source is viewed edge-on. This is because the emission is optically thick (see Sect. \ref{sec:caveats}) and hence the larger the emitting area the larger the emission.

The effect of viewing angle is considered by calculating the emission line for the fiducial envelope-only and envelope-plus-disk models and those with high mm opacity dust (i.e., large grains) with different viewing angles. Figure \ref{fig:angle} presents the integrated methanol flux for these models. This figure shows that the integrated flux only changes by a factor less than 2 when the viewing angle is changed. Therefore, for the rest of this work we consider a face-on view.

\subsubsection{Effects of envelope mass and luminosity}
\label{sec:mass_int}

Here we focus on comparing the total warm methanol mass and the (continuum subtracted) integrated methanol flux in various models.  The warm methanol mass is defined as the methanol mass inside the snow surface. More quantitatively, where methanol abundance is higher than 10$^{-9}$ in the envelope and higher than 10$^{-11}$ in the disk. Figure \ref{fig:mass_int} compares the warm methanol mass and its emission from different models with varying luminosity and envelope mass (or accretion rate).

Focusing on warm methanol mass, the general trend is that with increasing envelope mass and luminosity the warm methanol mass also increases. This is the same as what \cite{Nazari2022} found for the low-mass protostars. When the envelope mass increases the warm methanol mass increases simply because there is more mass. When the luminosity increases the warm methanol mass increases because the regions with temperatures above 68\,K (methanol sublimation temperature at the densities of our models) get larger. The slope of this relation with luminosity agrees well with the analytical formula of warm mass being proportional to $L^{3/4}$ (see gray solid line in Fig. \ref{fig:mass_int}; \citealt{Nazari2021}; \citealt{vanGelder2022}).

The warm methanol mass in Fig. \ref{fig:mass_int} is almost identical between the various models with the same luminosities and envelope masses (i.e., the models with or without a disk and those with large or small mm opacity dust). This is because the temperature structures are similar in most models with and without a disk and those with low or high mm opacity dust as explained in Sect. \ref{sec:gen_structure} (also see Fig. \ref{fig:temp}). Moreover, as shown in Fig. \ref{fig:meth_abund} there are almost no regions where methanol is photodissociated to decrease the warm methanol mass for the fiducial models and those with large mm opacity dust. This was different for the low-mass protostellar models with $M_{\rm E}$ of 1\,M$_{\odot}$ (\citealt{Nazari2022}) where larger photodissociation regions decreased the warm methanol mass.

There are slight variations (factor of $\lesssim 2$) between the warm methanol mass of models with small and large mm opacity dust grains or those with and without a disk (e.g. when $M_{\rm E}=50$\,M$_{\odot}$). The reason for such differences is the balance between various effects. Viscous heating in the disk becomes more effective if the accretion rate (or envelope mass) is larger. Therefore, it is expected to have colder disks for models with smaller envelope masses (and consequently lower accretion rates). For example for the lower end of envelope masses (i.e., 50\,M$_{\odot}$ or 150\,M$_{\odot}$), the envelope-plus-disk models are colder than the envelope-only models (see Figures \ref{fig:meth_abund_M50} and \ref{fig:meth_abund_M150}). Moreover, the depth of UV penetration in the envelope also affects the warm methanol mass. In Fig. \ref{fig:meth_abund_M50} (where $M_{\rm E}=50$\,M$_{\odot}$) large photodissociation regions are seen when the dust has a large mm opacity and a low UV opacity. This is especially important for lower envelope masses (i.e., lower densities), where it is easier for the UV to penetrate the envelope, photodissociate the methanol and decrease the warm methanol mass. 


\begin{figure*}
    \centering
    \includegraphics[width=0.8\textwidth]{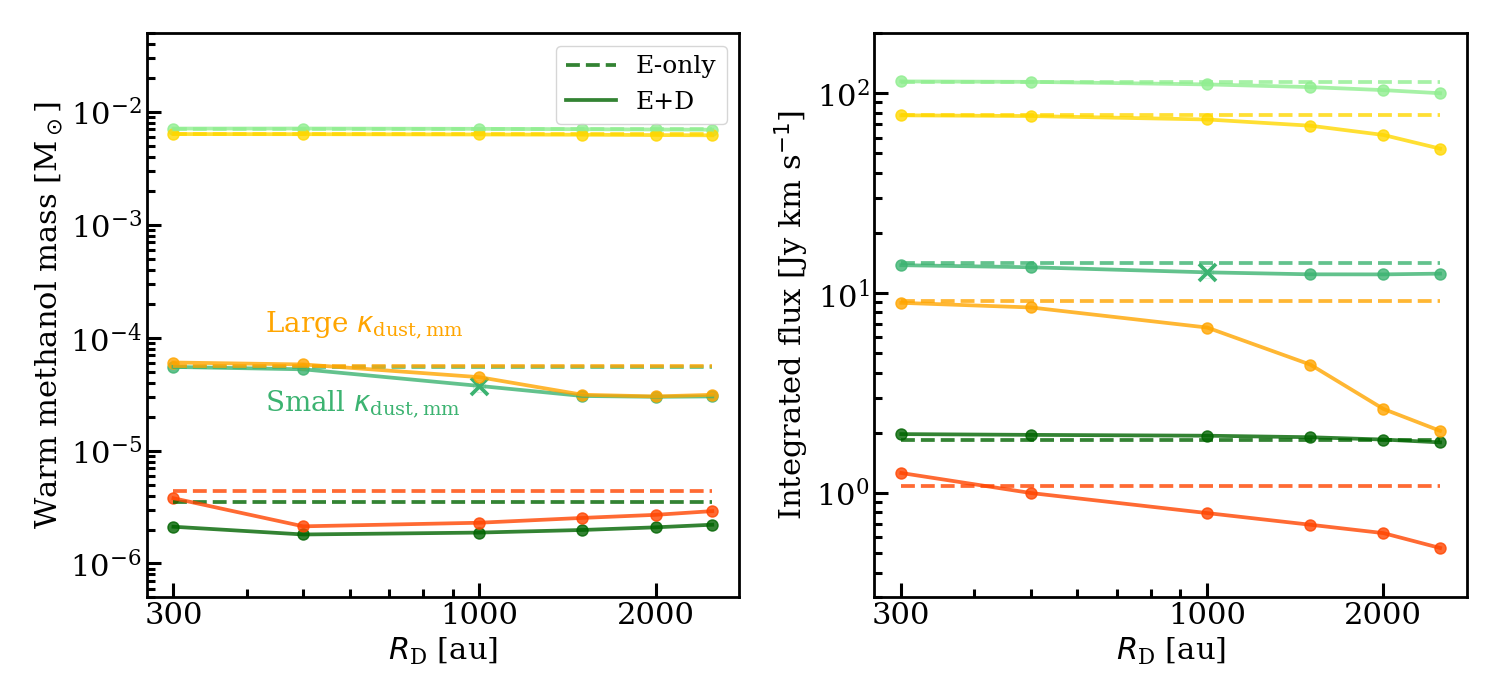}
    \caption{Warm methanol mass (left column) and integrated methanol emission (right column) for various disk sizes. The various shades of green and orange are used to indicate variations in luminosity. The luminosity from low to high is indicated by the darkest to lightest color. The models plotted here have luminosities $5\times 10^2$\,L$_{\odot}$, $1\times 10^4$\,L$_{\odot}$ and $5\times 10^6$\,L$_{\odot}$. The dashed lines present fiducial envelope-only models with various luminosities. The solid lines present the fiducial envelope-plus-disk models with various luminosities and disk radii. The shades of orange show models with large dust grains (high mm opacity) and the shades of green show those with small grains (low mm opacity).} 
    \label{fig:R_determine}
\end{figure*}

The general trends seen in the warm methanol mass (top row of Fig. \ref{fig:mass_int}) is reflected in the integrated continuum-subtracted methanol line fluxes (bottom row) especially for the trends seen with respect to luminosity. The integrated fluxes are mainly flat for various envelope masses while they increase with luminosity. Moreover, when the dust grains have a large mm opacity the integrated line fluxes are always smaller than when the grains have a small mm opacity (by factors of between about 2 and about 5) regardless of similar warm methanol masses (within factors of about 2) in most models. The reason for this is that when the grains have a high mm opacity, they can block the methanol emission in the envelope or hide it in the disk through the continuum over-subtraction effect (see Sect. 4.1 of \citealt{Nazari2022} for the explanation of this effect).

It is notable that for the luminosities, envelope masses and disk radii in Fig. \ref{fig:mass_int}, high mm opacity dust in the envelope and the continuum over-subtraction effect decrease the integrated fluxes by factors between about 2 and 5. However, it does not show a significant decrease (i.e., ${\sim}1$ order of magnitude) in methanol emission as was seen in low-mass protostars (\citealt{Nazari2022}). The difference between the effect of disk on methanol emission in low- and high-mass protostars is presented in Fig. \ref{fig:low_high_comp}. This figure shows that the methanol emission for the models with disk and those without one are similar for high-mass protostars in this work. More quantitatively, the ratio of emission between the two models is between ${\sim}0.6$ and ${\sim}1$ (less than a factor 2 difference). In contrast, the methanol emission for the low-mass models can be dropped by a factor of ${\sim}10$ if disk and high mm opacity dust are included (\citealt{Nazari2022}).

Finally, it is important to note that the methanol emission is optically thick (see Sect. \ref{sec:caveats}). This can also be deduced by comparing the variations in warm methanol and those in integrated flux. The warm methanol mass varies by ${\sim} 3-4$ orders of magnitude as a function of envelope mass and luminosity. While the integrated flux spans a range of ${\sim}2$ orders of magnitude as a function of luminosity and only a range of factor ${\sim}2$ as a function of envelope mass. Apart from the fact that warm methanol mass increases as a function of luminosity, the reason that an increase is seen in integrated flux for the models with more optically thick methanol lines is the larger area of the emission. If the line is optically thick methanol emission would be proportional to the emitting area and the larger the luminosity the larger the emitting area of methanol (see Fig. \ref{fig:temp_grid} for the temperature structure of various models).

\subsubsection{Effects of disk size}
\label{sec:disk_size}

Figure \ref{fig:R_determine} presents the variation of warm methanol mass and its emission with disk radius for three different luminosities. The warm methanol mass in all models is constant and is not a function of disk size. Moreover, the warm methanol mass is similar between the models with and without disk. This is because of the similar temperature structures explained in Sect. \ref{sec:gen_structure}.

The methanol emission also does not show a relation with disk radius when the dust has a low mm opacity. There is a factor of at most two between the envelope-only models with high and low mm opacity dust. However, when the dust has a high mm opacity the emission decreases with increasing disk size. Large disks cause a large (factor of at most ${\sim}5$) drop in integrated flux of the envelope-plus-disk model compared with the envelope-only model with high mm opacity dust.

A disk with a minimum radius of ${\sim}1000$\,au and large mm opacity dust is necessary for a drop of at least a factor of about two in methanol emission compared with the envelope-only and envelope-plus-disk models with small mm opacity dust (at $L=10^4$\,L$_{\odot}$). Moreover, for a drop of an order of magnitude, a disk size of ${\sim}2000-2500$\,au with high mm opacity dust is needed. These large drops are due to the continuum over-subtraction effect in the disk (see Fig. \ref{fig:cont_over_sub_2000au}). In addition, the radius at which a drop is seen in methanol emission in the disk-plus-envelope models increases with luminosity. In other words, larger disk sizes are needed for a large decrease in methanol emission if a source has a high luminosity.

\section{Discussion}
\label{sec:discussion}




\subsection{Comparison with observations}
\label{sec:obs}

The main goal of this work is to examine whether it is possible to explain the spread in observations of methanol emission discussed in \cite{vanGelder2022}. Therefore, in this section we compare the integrated flux of methanol from the models with that of the same methanol line in ALMAGAL observations (CH$_3$OH 4$_{2,3,1}$-3$_{1,2,1}$, $\nu = 218.4401$\,GHz, $E_{\rm up}=45.5$\,K and $A_{\rm i,j} = 4.7 \times 10^{-5}$\,s$^{-1}$) from \cite{vanGelder2022}.

\begin{figure*}
    \centering
    \includegraphics[width=15cm]{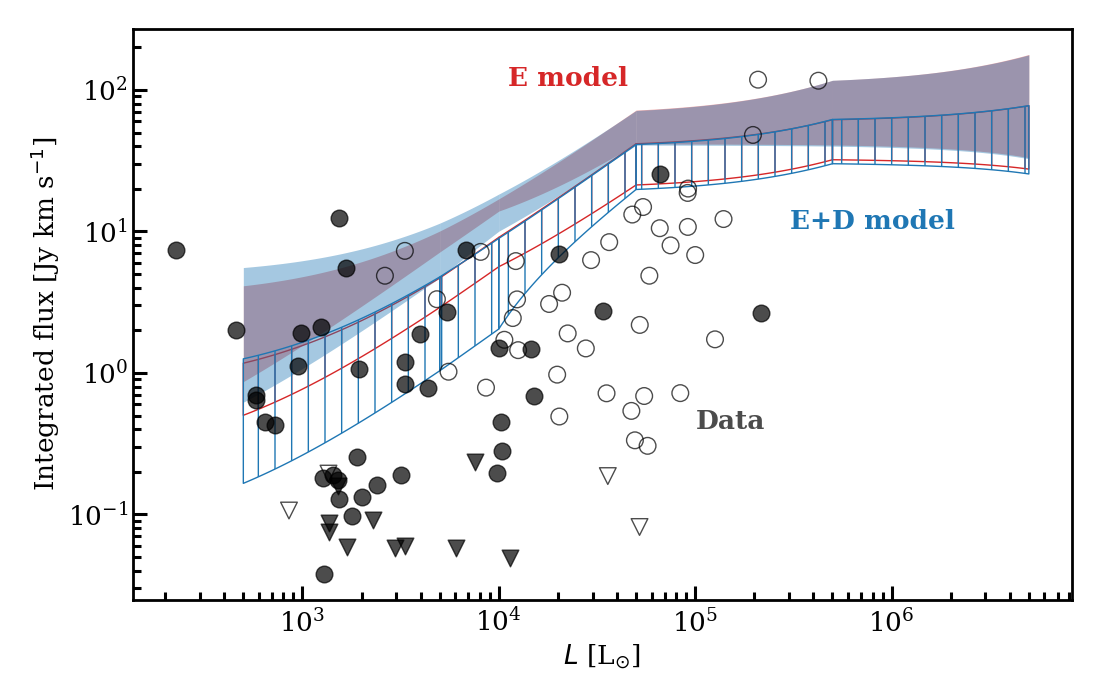}
    \caption{The comparison of models with observations of ALMAGAL sources. The same methanol line is used in both the models and the observations. Moreover, the integrated fluxes from the observations are normalized to a distance of 4\,kpc to match those from the models. The black data points present the observations where the circles are detections and triangles are upper limits. The empty symbols indicate sources that have their $L/M$ from \cite{Elia2017} above 22.4\,L$_{\odot}$\,M$_{\odot}^{-1}$ as proposed `HII region candidates'. The smooth red and blue regions show the results from the envelope-only and envelope-plus-disk models with small mm opacity dust grains. The striped regions show the same for models with large mm opacity dust. The regions where the models fall (blue and red) are found by simply connecting the integrated fluxes at the six different luminosities considered in this work in the linear space.} 
    \label{fig:obs}
\end{figure*}

Figure \ref{fig:obs} presents the comparison of our models with observations. First, the scaling of flux with luminosity in our models and the data is apparent as also explained in Sect. \ref{sec:mass_int}. Second, the regions indicating envelope-only and envelope-plus-disk models with small mm opacity dust grains (red and blue smooth regions) coincide. This is expected from the similar temperature structures and warm methanol masses between the two models as discussed in Sections \ref{sec:gen_structure} and \ref{sec:warm_mass}. Third, the integrated fluxes from the two models when the grains have a large mm opacity are also similar, with the envelope-plus-disk models having around a factor of about two to three lower integrated fluxes when the luminosities are below around $10^4$\,L$_{\odot}$ due to continuum over-subtraction. Finally, the models cannot explain the whole range (${\sim} 2$ orders of magnitude) of methanol emission. Although they fail to match the observations with integrated methanol fluxes below ${\sim}0.1$\,Jy\,km\,s$^{-1}$, they do explain the data better when the luminosities are lower. The models especially miss the data points at higher luminosities ($L\simeq 10^4-10^5$\,L$_{\odot}$).

Therefore, disks and dust optical depth effects are not as effective in massive protostars to decrease the methanol emission compared with low-mass protostars studied in \cite{Nazari2022} where disks could explain the spread in observations well. Although they can explain almost two order of magnitude spread in methanol emission at low luminosities (${\sim} 5 \times 10^2 \sim 10^4$\,L$_{\odot}$), they cannot explain the data at higher luminosities. Therefore, other effects are needed which are discussed further below.

\subsection{Alternative explanations}
\subsubsection{Larger disk sizes and lower envelope masses}
\label{sec:disk_stability}

One way to further lower the methanol emission is to increase the disk radii in our models (see Fig. \ref{fig:R_determine}). This is not realistic due to disks becoming more unstable as they become larger and more massive.

The disks considered here are stable by definition from the calculation of Toomre $Q$ parameter. We have calculated the Toomre Q parameter for our disks but because the disk masses and the disk radii are changed such that $M_{\rm D}/R_{\rm D}^2$ always stays equal to 0.003\,M$_{\odot}$\,R$_{\odot}^{-2}$, by definition our disks are always stable. However, the maximum disk radius of 2500\,au in our models is the most extreme limit on the disk radius in massive protostars from observations (e.g., see \citealt{Jimenez2012}; \citealt{Hunter2014}; \citealt{Zapata2015}; \citealt{Williams2022}). We especially note that it is easier to observe the larger and more massive disks, hence if disks larger than ${\sim}2500$\,au have not yet been observed this could be an indication that they do not exist.

In fact many of the observations of large disks or rotating structures (referred to as `toroids' see \citealt{Beltran2016}) show evidence for fragmentation once they are observed with higher angular resolution (e.g. \citealt{Beuther2009}; \citealt{Beuther2017}; \citealt{Ilee2016}; \citealt{Ilee2018}; \citealt{Beuther2018}; \citealt{Suri2021}). One of the best studied large Keplerian disks known to date is that around the protostar AFGL 4176 which was found to have a radius of 2000\,au (\citealt{Johnston2015}; also see \citealt{Bogelund2019} for extent of emission from various species). Recently \cite{Johnston2020} used even higher angular resolution data of this disk to calculate the Toomre Q parameter. They concluded that the outer part of the disk is unstable and is prone to fragmentation. Therefore, disks larger than 2500\,au are not realistic and one cannot simply increase the disk radius to explain the whole range of methanol emission observed in Fig. \ref{fig:obs}.

Another parameter that can be changed to decrease the methanol emission is envelope mass (see Fig. \ref{fig:mass_int}). This is because the lower the envelope mass, the lower the warm methanol mass and hence the lower methanol emission. Especially since the emission will become more optically thin toward this end. The ranges of envelope masses observed for high-mass protostellar systems especially those shown in Fig. \ref{fig:obs} are mostly above 50\,M$_{\odot}$ (\citealt{vanderTak2000}; \citealt{Schuller2009}; \citealt{Dunham2011}; \citealt{Elia2017}; \citealt{Konig2017}). Therefore, lowering the mass is not a realistic solution to decrease the methanol emission. 


\begin{figure}
  \resizebox{0.6\hsize}{!}{\includegraphics{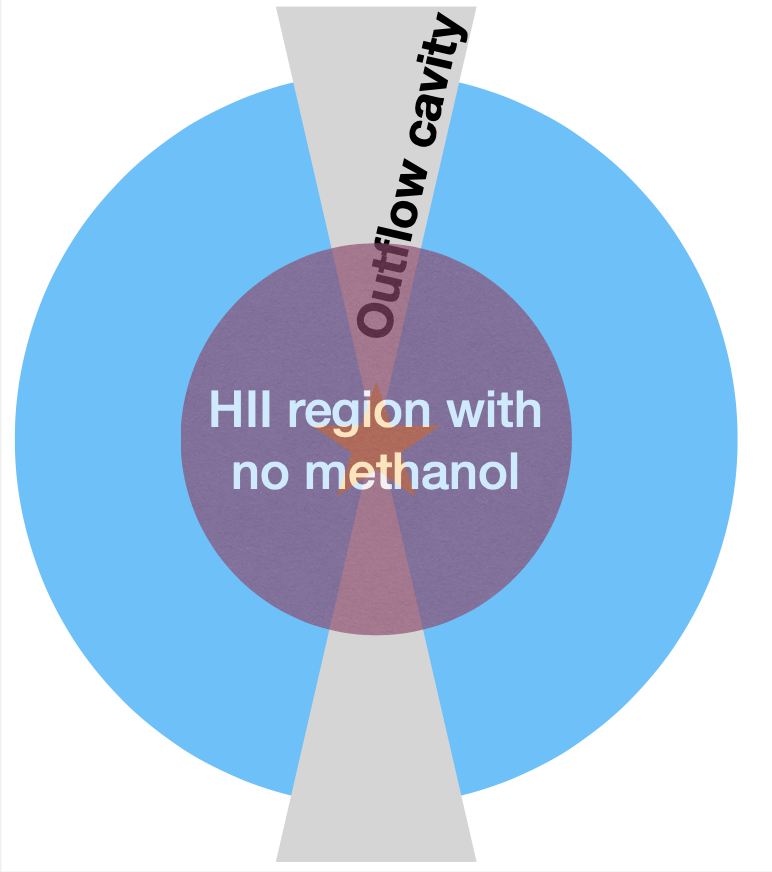}}
  \centering
  \caption{Sketch of the potential hypercompact/ultracompact HII region around a high-mass protostar. It is expected that methanol is absent in this region as the gas is atomic and ionized.}
  \label{fig:HII_sketch}
\end{figure}

\subsubsection{Absence of methanol}
\label{sec:HII}



This section considers the case where the abundance of methanol in some high-mass systems is intrinsically lower. One way to have less methanol is to have large HII regions where methanol is absent. As explained in Sect. \ref{sec:phycical_structure} HII regions are not included in our models. A self-consistent modeling of the HII region, including its extent is beyond the scope of the paper. However, it is expected that a star with blackbody radiation at $T_{\star} = 40000$\,K has ${\sim} 40\%$ of its emitted photons with energies larger than the energy needed to ionize hydrogen.

Here we explore how the methanol emission would change if we include spheres with various radii where methanol is absent in our fiducial envelope-only model (see Fig. \ref{fig:HII_sketch}). These spheres are to mimic the effect of a potential HC/UC HII region present in a protostellar system. We assume no methanol in the HII regions as by definition the gas is atomic and ionized in these regions. The radii considered for the spheres with no methanol inside are 50\,au, 200\,au, 500\,au, 1000\,au, 5000\,au and 10000\,au. The values assumed here are in line with the extents suggested by modeling and observational works for HC/UC HII regions (\citealt{Keto2003}; \citealt{Sewilo2004}; \citealt{Hoare2007}; \citealt{Cyganowski2011}; \citealt{Sanchez2013UCHII}; \citealt{Ilee2016}; \citealt{Williams2022}). We note that sources with disks can also have an HII region related to the disk wind (\citealt{Hollenbach1994}), however modeling such disk winds is beyond the scope of this paper. Thus the effect of HII regions on methanol emission is only considered in the envelope-only models which have similar methanol fluxes to those models with disks (see Fig. \ref{fig:obs}). 

\begin{SCfigure*}
    \centering
    \includegraphics[width=13cm]{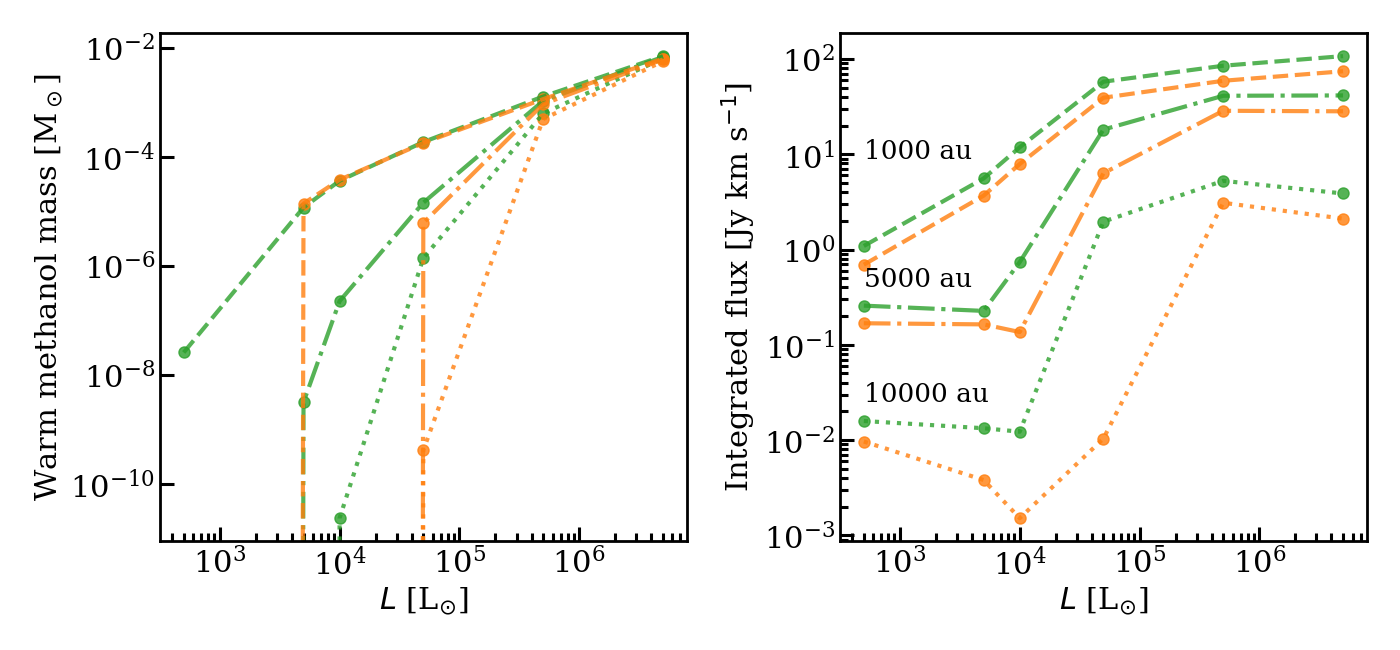}
    \caption{The warm methanol mass (left) and integrated line flux of methanol (right) for envelope-only fiducial models with different luminosities and carved HC/UC HII regions. Green and orange show low and high mm opacity dust models. Dashed lines show when the size of HC/UC HII region is 1000\,au, dashed dotted lines show the same for 5000\,au and the dotted lines show the same for 10000\,au. An HII region of size ${\sim}10000$\,au is needed for a drop of $\gtrsim 2$ orders of magnitude in methanol emission.} 
    \label{fig:HII}
\end{SCfigure*}


Figure \ref{fig:HII} presents the warm methanol mass and integrated line fluxes for the fiducial envelope-only models with different assumed HC/UC HII region sizes above 500\,au (those with assumed HII region sizes below or equal to 500\,au are shown in Fig. \ref{fig:HII_small}). The warm methanol mass does not change significantly compared with the fiducial envelope-only model (see Fig. \ref{fig:mass_int}) for most models with $L \gtrsim 10^4$\,L$_{\odot}$ and HII region size of 1000\,au. Moreover, the integrated methanol flux does not change significantly compared with the fiducial envelope-only model for all luminosities when the carved HII region is $\leq 1000$\,au (also see Fig. \ref{fig:HII_small}). However, when the carved region has a size of $\geq 5000$\,au the warm methanol mass and methanol emission drop. More quantitatively, the methanol emission is decreased by $\sim 1$ order of magnitude when the carved region is 5000\,au and the the luminosities are $\lesssim 10^4$\,L$_{\odot}$. The integrated flux decreases even more (by
$\gtrsim 2$ orders of magnitude) for all luminosities when the HII region is 10000\,au. This shows that to have a large drop in methanol emission such that the models match the data in Fig. \ref{fig:obs}, an HII region of size of $ > 5000$\,au is needed. These sizes would fall in the category of a large HC HII region or an UC HII region. 

\cite{Elia2017} categorize the sources with $L/M > 22.4$\,L$_{\odot}$\,M$_{\odot}^{-1}$ as those with `HII region candidates' where HII region here mainly means ultracompact or compact HII regions (also see \citealt{Cesaroni2015}). Therefore, the sources that satisfy this criterion based on the luminosities and masses given by \cite{Elia2017} for these sources are highlighted in Fig. \ref{fig:obs} by empty symbols. It is interesting that almost all the sources with luminosities between 10$^4$\,L$_{\odot}$ and 10$^5$\,L$_{\odot}$ could be potential HII region candidates. Hence, the data in that part of the plot can be explained by those sources having an UC or compact HII region. We note that if a source hosts a hypercompact/ultracompact HII region, it is not guaranteed that the methanol emission is low. This only happens if the HII region is large enough compared with the methanol sublimation region (also see Fig. \ref{fig:HII}). However, if methanol emission is low, it is important to consider the potential effects from a hypercomapct/ultracompact HII region.

It is not clear why some sources with luminosities between 10$^3$\,L$_{\odot}$ and 10$^4$\,L$_{\odot}$ have lower methanol emission than our models. Another way to decrease the methanol abundance in a protostar could be its destruction by X-rays. It has recently been found that X-rays can cause lower methanol emission in low-mass protostars (\citealt{Notsu2021}). In particular they find that for $L_{\rm X} \gtrsim 10^{30}-10^{31}$\,erg\,s$^{-1}$ the methanol abundance decreases significantly. \cite{Stauber2005} consider X-ray chemistry for high-mass protostars and find an X-ray luminosity $\gtrsim 10^{31}$\,erg\,s$^{-1}$ for the high-mass source AFGL 2591. Based on these results and those of \cite{Notsu2021} methanol in high-mass protostars similar to AFGL 2591 could be destroyed by X-ray chemistry in the envelope. However, whether such a phenomenon is important is still to be confirmed. Especially given that \cite{Benz2016} found no evidence of X-ray chemistry in a sample of low- and high-mass protostars on scales of ${\sim}1000$\,au.

If the study of \cite{Notsu2021} can be applied to high-mass protostars, it is expected that HCO$^{+}$ is abundant on-source when methanol is not detected or its flux is low. That is because X-rays also destroy water (\citealt{Notsu2021}) and where water is absent HCO$^+$ is abundant (\citealt{vantHoff2021}). Therefore, to solve the mystery of low methanol emission in the massive protostars with luminosities between 10$^3$\,L$_{\odot}$ and 10$^4$\,L$_{\odot}$ in Fig. \ref{fig:obs}, high spatial resolution studies with deep observations of HCO$^{+}$ and its isotopologues in these sources are needed. 

Finally, it is also possible that methanol simply forms less efficiently toward massive protostars because of their potentially warmer pre-stellar phase. This would agree with observations of \cite{vanGelder2022b}, who measured lower D/H ratios toward massive protostars compared with their low-mass counterparts. They interpreted their results as either a warmer pre-stellar phase or shorter pre-stellar lifetimes for these massive sources compared with low-mass protostars. However, if this is the case one would expect that it would happen for all the sources in Fig. \ref{fig:obs} and affect them all similarly. Therefore, the reason for low methanol emission of some sources is probably not the low production rate of this molecule.

\subsection{Caveats}
\label{sec:caveats}

One important fact about the methanol emission in our models is that it is optically thick. This was already pointed out in Sect. \ref{sec:mass_int} where the integrated methanol emission spans a smaller range than the warm methanol mass (Fig. \ref{fig:mass_int}). This can be confirmed by calculating the line optical depth in the fiducial models. Figure \ref{fig:tau_line} presents a radial cut through the line optical depth in the fiducial envelope-only and envelope-plus-disk models. This figure shows that the emission is optically thick inside the methanol snow surface for these two models. 

Given that the line is already optically thick (the emission is proportional to the emitting area), increasing the abundance of methanol in the inner and outer disk by two orders of magnitude (based on the findings of \citealt{Bogelund2019}) should not change the integrated emission significantly. Therefore, we specifically test this for the fiducial envelope-plus-disk model. The line emissions are shown in Fig. \ref{fig:high_abund}. The integrated flux is only $<1\%$ larger when the abundance is higher. Therefore, the conclusions made here should not change if higher methanol abundances are assumed in the disk as long as optically thick methanol lines are considered. Moreover, we test for a case where the methanol abundance in the disk is one order of magnitude lower than assumed in this work. Again the integrated methanol flux is only $<1\%$ smaller when the methanol abundance is lower. Therefore, this cannot be the reason for the low methanol emission in the observations.

Another assumption is the protostellar mass and temperature. A few models with stellar temperature of 20000\,K and stellar mass of 10\,M$_{\odot}$ were run to test the effect. However, the change in the integrated methanol flux was less than a factor ${\sim}1.5$.

In our models we do not include the effect of shocks in enhancing the methanol emission. Studies show that shocks can enhance the abundance of various molecules including methanol (\citealt{Csengeri2019}; \citealt{vanGelder2021}; \citealt{Garufi2022}). Therefore, including this effect of shocks would increase the methanol emission and separate the models from observations in Fig. \ref{fig:obs} even more. 

Finally, we have made an assumption that the hydrogen ionising photons from the protostar are re-emitted at a longer wavelength (assumed to be Lyman-$\alpha$ here) before interacting with the dust. In reality the photons can be emitted at longer wavelengths via a forest of lines from atomic and ionized species. Hence, we consider a case where these ionising photons are completely eliminated from the system representing a lower limit on methanol emission. The reality is more similar to the main grid run in this paper in terms of including these photons. When the ionising photons are deleted in the fiducial envelope-only and envelope-plus-disk models the integrated methanol emission is only a factor of $\lesssim 2$ smaller than the models considered here. Therefore, the large spread seen in the data cannot be explained by a change in the exact spectrum emerging from the HII region surrounding the protostar.

\section{Conclusions}
\label{sec:conclusions}

In this work we considered the importance of disks in decreasing methanol emission in high-mass protostars. We studied two models: an envelope-only model and an envelope-plus-disk model. Both models include low and high mm opacity dust grains separately (representing small and large grains). A large range of parameters were considered in envelope-only and envelope-plus disk models. The luminosities range from $5\times 10^2$\,L$_{\odot}$ to $5\times 10^6$\,L$_{\odot}$, envelope masses from 50\,M$_{\odot}$ to 1000\,M$_{\odot}$ and disk radii from 300\,au to 2500\,au. Our conclusions are summarized below:

\begin{itemize}
    \item The temperature structures of high-mass protostellar systems with and without a disk are similar. This is because the disk mid-plane is hot due to viscous heating in the disk and disk shadowing is not as effective as it is for low-mass protostellar disks. Moreover, the temperature structures of models with low and high mm opacity are also similar. The warm methanol mass is hence similar in these models due to the similar temperature structures.
    \item Dust with large mm opacity blocks the methanol emission in the envelope and hides it in the disk through continuum over-subtraction effect. The minimum disk size to observe a factor of two drop between the envelope-only models with small grains (low mm opacity) and the envelope-plus-disk models with large grains (high mm opacity) increases with luminosity. At $L=10^4$\,L$_{\odot}$ this disk size is ${\sim}1000$\,au. For an order of magnitude drop in emission at $L=10^4$\,L$_{\odot}$ a minimum disk size of ${\sim}2000-2500$\,au is needed.
    \item The temperature inversion effect that was suggested by \cite{Barr2020} in the disk to explain the absorption lines toward two massive protostars is indeed found in our models at 50\,au but only in models with large mm opacity dust. This effect is only observed when the envelope mass is ${\gtrsim} 550$\,M$_{\odot}$ or the accretion rate is ${\gtrsim} 3.6 \times 10^{-3}$\,M$_{\odot}$\,yr$^{-1}$. 
    \item The entire spread in observed methanol emission toward high-mass protostars (especially sources with high luminosities larger than ${\sim} 10^4$\,L$_{\odot}$) cannot be explained by the presence of a disk or dust opacity. This is in contrast with models by \cite{Nazari2022} for low-mass protostars. A possible explanation for low methanol emission of sources with high luminosities could be that they host a HC/UC HII region as also suggested by their $L/M$ ratio. 
\end{itemize}

The lowest methanol emission in low-luminosity objects ($L \simeq 10^3-10^4$\,L$_{\odot}$ ) might be due to destruction of methanol by X-rays in those sources. Hence, these object are prime targets to study X-ray chemistry. A future step is to study these sources with deep and higher angular resolution observations.


\begin{acknowledgements}
We thank the referee for the constructive comments. We thank E. F. van Dishoeck for the constructive comments and discussions. We also thank M. L. van Gelder for the constrictive comments and providing us with the integrated fluxes in the ALMAGAL sources. This work is supported by grant 618.000.001 from the Dutch Research Council (NWO). Support by the Danish National Research Foundation through the Center of Excellence “InterCat” (Grant agreement no.: DNRF150) is also acknowledged. B.T. is a Laureate of the Paris Region fellowship program, which is supported by the Ile-de-France Region and has received funding under the Horizon 2020 innovation framework program and Marie Sklodowska-Curie grant agreement No. 945298. G.R. acknowledges support from the Netherlands Organisation for Scientific Research (NWO, program number 016.Veni.192.233) and from an STFC Ernest Rutherford Fellowship (grant number ST/T003855/1). This project has received funding from the European Research Council (ERC) under the European Union's Horizon Europe Research \& Innovation Programme under grant agreement No 101039651 (DiscEvol). Views and opinions expressed are however those of the author(s) only and do not necessarily reflect those of the European Union or the European Research Council Executive Agency. Neither the European Union nor the granting authority can be held responsible for them.
\end{acknowledgements}

%
%

\bibliographystyle{aa}
\bibliography{high_mass_methanol}

\begin{appendix}

\section{Passive heating vs viscous heating}
\label{sec:pass_visc}

The analytical solution for the temperature in the mid-plane from passive heating can be given by (\citealt{Chiang1997}; \citealt{Dullemond2001}; \citealt{Dullemond2018}) 

\begin{equation}
    T_{\rm mid, irr} = \left(\frac{0.5 \varphi L}{4 \pi R^2 \sigma_{\rm SB}} \right)^{1/4},
    \label{eq:T_irr}
\end{equation}

\noindent where, $\sigma_{\rm SB}$ is the Stefan–Boltzmann constant and $\varphi$ is the flaring angle. Here, $\varphi$ is assumed as 0.2 to match the temperatures found from RADMC-3D models in the outer radii (see the left panel of Fig. \ref{fig:temp_compare}). This equation is a result of balancing heating from the star and cooling. The heating from the star depends very much on the geometry and the angle that the radiation impinges on the disk.

Knowing that the dissipated power per unit area due to viscosity is given by $\sigma_{\rm SB} T_{\rm surf, visc}^4$, the temperature that this dissipated energy corresponds to is (\citealt{Lodato2008})

\begin{equation}
    2\sigma_{\rm SB} T_{\rm surf, visc}(R)^4 = \int_{-\infty}^{\infty}{Q(R,z)}dz = \frac{3 \dot{M}}{4\pi} \frac{GM_{\star}}{R^3} \left(1-\sqrt{\frac{R_{\rm in}}{R}}\right).
    \label{eq:T_visc}
\end{equation}

\begin{figure}
  \resizebox{0.9\hsize}{!}{\includegraphics{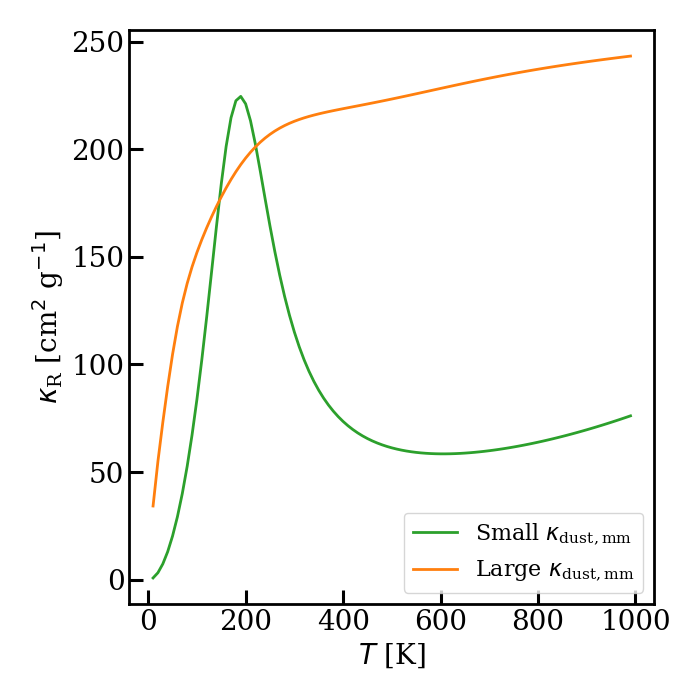}}
  \caption{The Rosseland mean opacity as a function of temperature.}
  \label{fig:rosseland_opac}
\end{figure}

\noindent Factor 2 on the left hand side of this equation comes from the fact that a disk has two sides. This is the analytic approximation of disk surface temperature from viscous heating. However, as explained in Sect. \ref{sec:gen_structure} the mid-plane temperature is hotter than the surface layers due to the higher densities and hence higher optical depth in the mid-plane. The mid-plane temperature can be approximated by (\citealt{Armitage2010})

\begin{equation}
    T_{\rm mid,\, visc}^{4} \simeq \frac{3}{4} \tau T_{\rm surf, visc}^{4},
    \label{eq:T_mid_visc}
\end{equation}

\noindent where $\tau = \kappa_{\rm R} \Sigma_{\rm dust} /2$ and $\kappa_{\rm R}$ is the Rosseland mean opacity. The Rosseland mean opacity can be calculated using 

\begin{equation}
    \frac{1}{\kappa_{\rm R}} = \frac{\int_{0}^{\nu^\prime} \frac{1}{\kappa_{\nu}} \frac{dB_{\nu}(T)}{dT} d\nu}{\int_{0}^{\nu^\prime} \frac{dB_{\nu}(T)}{dT} d\nu},
    \label{eq:rosseland_opac}
\end{equation}

\noindent where, $\kappa_{\nu}$ is the absorption opacity and $B_{\nu}(T)$ is the Planck function. The integrals are calculated for $\nu$ going from zero to $\nu^\prime$, where $\nu^\prime$ is assumed to be the frequency of the photons that can ionize hydrogen (the wavelength of these photons would be ${\sim} 0.1$\,$\mu$m). Equation \eqref{eq:rosseland_opac} is dependent on the temperature, so to calculate $\kappa_{\rm R}$ an initial temperature of $T_{\rm surf, visc}$ from Eq. \eqref{eq:T_visc} is assumed to give an initial value of $\kappa_{\rm R}$. This initial value is then used to find the temperature from Eq. \ref{eq:T_mid_visc}. This procedure is done iteratively until $\kappa_{\rm R}$ from one iteration to the other varies by less than 0.01. Figure \ref{fig:rosseland_opac} shows the resulting $\kappa_{\rm R}$ as a function of temperature. In reality, the total heating in the disk mid-plane ($\propto T_{\rm mid, total}^4$) is the sum of that from viscosity and radiation from the star and hence

\begin{equation}
    T_{\rm mid, total}^{4} \simeq \frac{3}{4} \tau T_{\rm surf, visc}^{4} + T_{\rm mid, irr}^{4}.
    \label{eq:T_mid}
\end{equation}

In a similar way the disk surface temperature can be derived. The disk surface temperature ($T_{\rm surf, visc}$) due to viscosity can be found from Eq. \eqref{eq:T_visc}. The temperature in the disk surface due to passive heating is given by (\citealt{Dullemond2001})

\begin{equation}
    T_{\rm surf, irr} = \left(\frac{L}{\epsilon 8 \pi \sigma_{\rm SB} R^2}\right)^{1/4},
    \label{eq:T_surf_irr}
\end{equation}

\noindent where $\epsilon = \kappa_{\rm P} (T_{\rm surf, irr})/\kappa_{\rm P} (T_{\star})$. Moreover, $\kappa_{\rm P}$ is the Planck mean opacity given by

\begin{equation}
    \kappa_{\rm P} = \frac{\int_{0}^{\nu^\prime} \kappa_{\nu} B_{\nu}(T) d\nu}{\int_{0}^{\nu^\prime} B_{\nu}(T) d\nu}.
    \label{eq:planck_opac}
\end{equation}

\noindent Planck mean opacity similar to the Rosseland mean opacity is dependant on the temperature. Therefore, to calculate $\kappa_{\rm P} (T_{\rm surf, irr})$, first mid-plane temperature due to passive heating ($T_{\rm mid, irr}$) is used to find $\kappa_{\rm P}$ and then that is used in Eq. \eqref{eq:T_surf_irr} to find $T_{\rm surf, irr}$ to be used again for calculation of $\kappa_{\rm P}$. This process is done iteratively until the value of $\kappa_{\rm P}$ converges.

Finally, the heating in the disk surface ($\propto T_{\rm surf, total}^4$) is the sum of that due to viscous heating and passive heating, in other words it is given by

\begin{equation}
    T_{\rm surf, total}^{4} \simeq T_{\rm surf, visc}^{4} + T_{\rm surf, irr}^{4}.
    \label{eq:T_surf}
\end{equation}

\section{Additional plots}

\begin{figure*}
    \centering
    \includegraphics[width=0.9\textwidth]{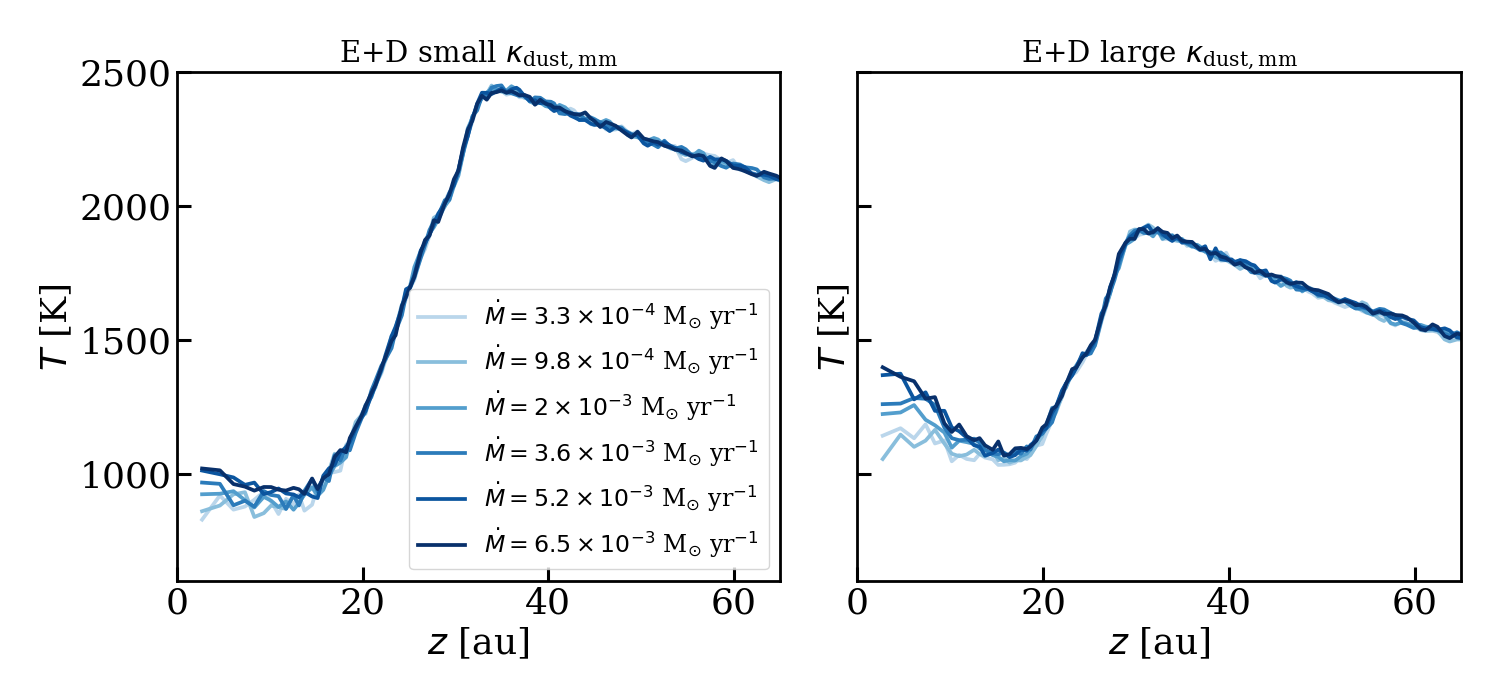}
    \caption{The same as Fig. \ref{fig:temp_vertical} but for when the bolometric luminosity is $5 \times 10^{5}$\,L$_{\odot}$ and the vertical temperature cut is made at 30\,au.} 
    \label{fig:temp_vertical_L5e5}
\end{figure*}
\begin{figure*}
    \centering
    \includegraphics[width=\textwidth]{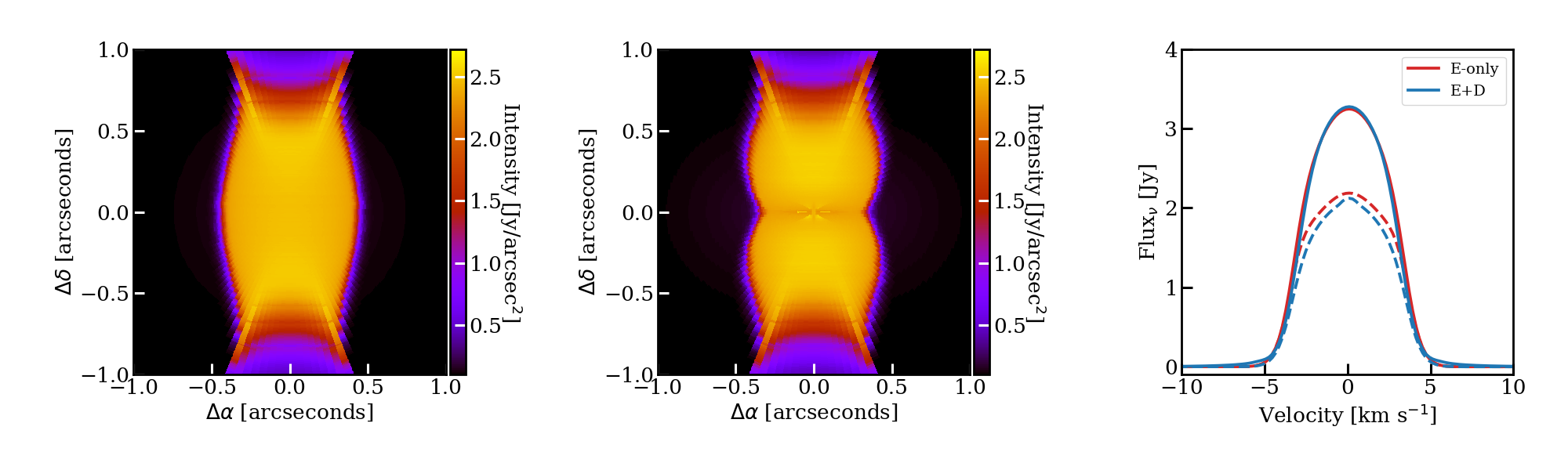}
    \caption{Methanol emission from the fiducial envelope-only and envelope-plus-disk models. The left and middle panel show the emission at the peak of the line viewed edge-on for the two models with no dust included in these two particular models so that the methanol emission can be seen without optical depth effects from the dust (dust is included in all other models unless otherwise stated). The right panel shows the continuum subtracted line flux at an assumed source distance of 4\,kpc when viewed edge-on (solid lines) and face-on (dashed lines).}
    \label{fig:meth_emission}
\end{figure*}
\begin{figure}
  \resizebox{\hsize}{!}{\includegraphics{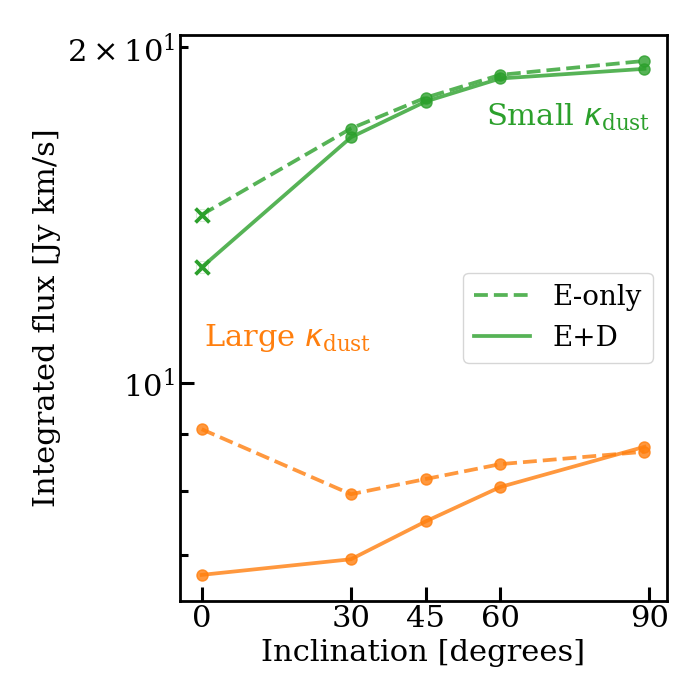}}
  \centering
  \caption{The integrated methanol flux as a function of viewing angle for the fiducial envelope-only and envelope-plus-disk models and those with high mm opacity dust. There is a factor less than 2 difference in the integrated flux when the viewing angle changes.}
  \label{fig:angle}
\end{figure}
\begin{figure*}
    \centering
    \includegraphics[width=0.8\textwidth]{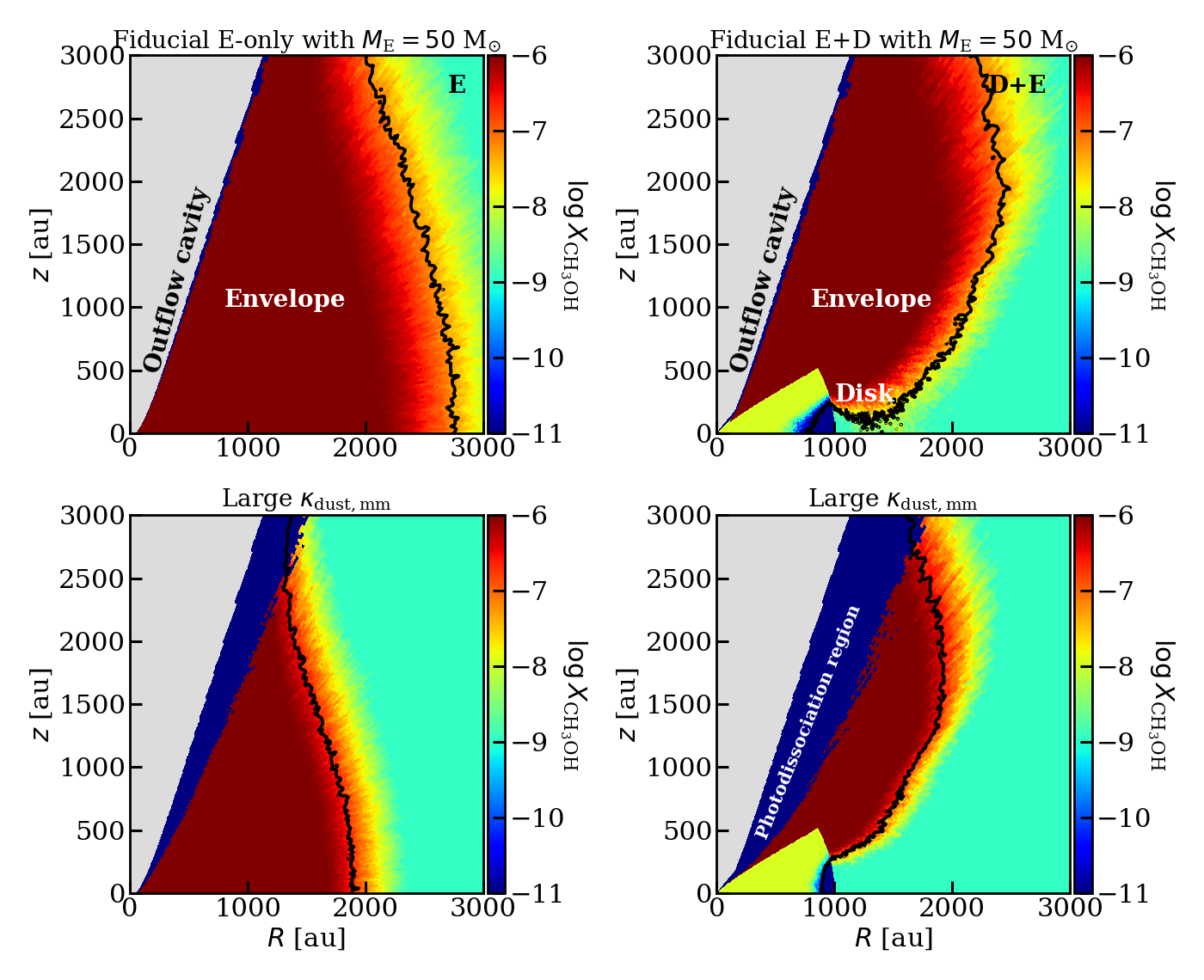}
    \caption{The same as Fig. \ref{fig:meth_abund} but now for $M_{\rm E}=50$\,M$_{\odot}$. The photodissociation regions for the models with large mm opacity dust are significant.} 
    \label{fig:meth_abund_M50}
\end{figure*}
\begin{figure*}
    \centering
    \includegraphics[width=0.8\textwidth]{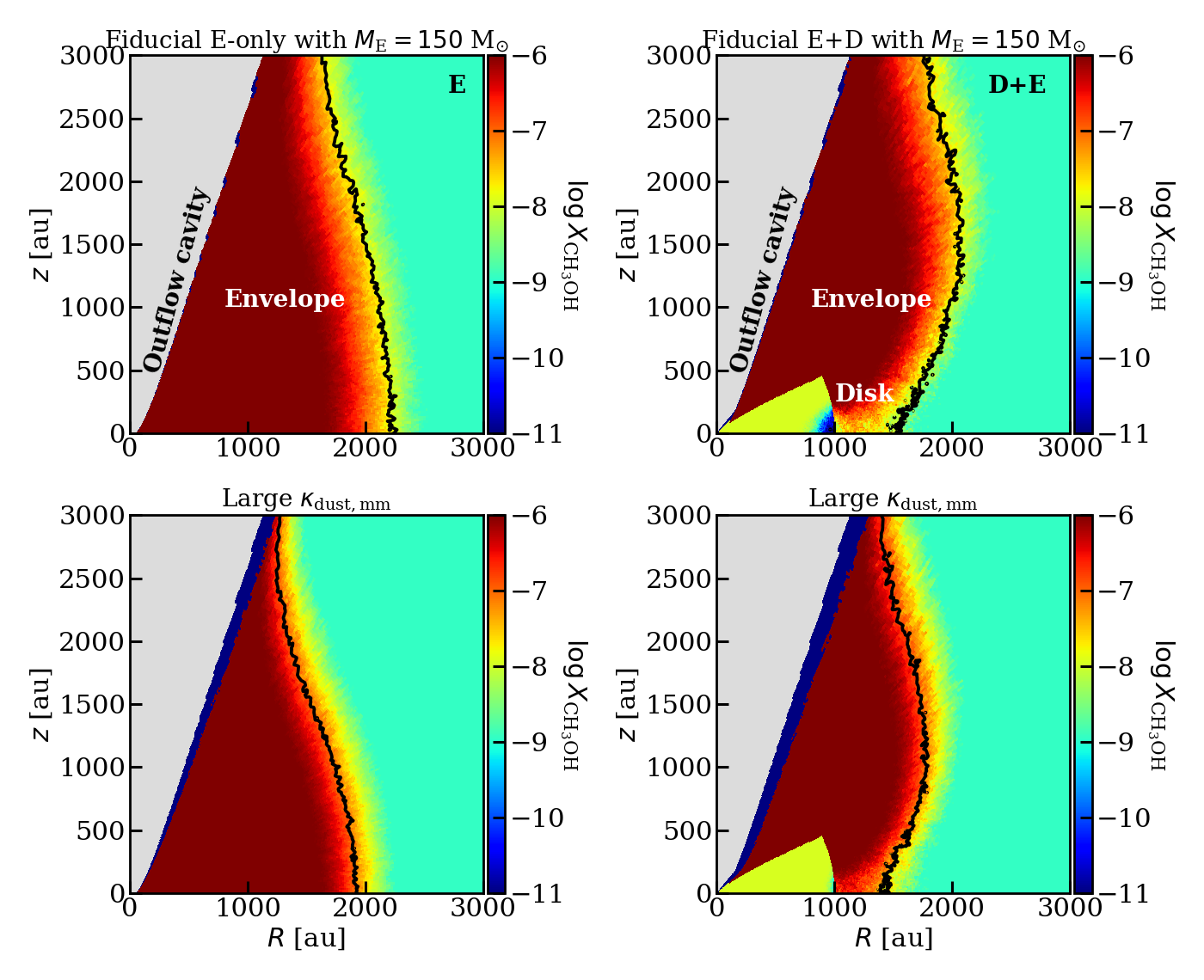}
    \caption{The same as Fig. \ref{fig:meth_abund} but now for $M_{\rm E}=150$\,M$_{\odot}$.} 
    \label{fig:meth_abund_M150}
\end{figure*}
\begin{figure*}
    \centering
    \includegraphics[width=0.8\textwidth]{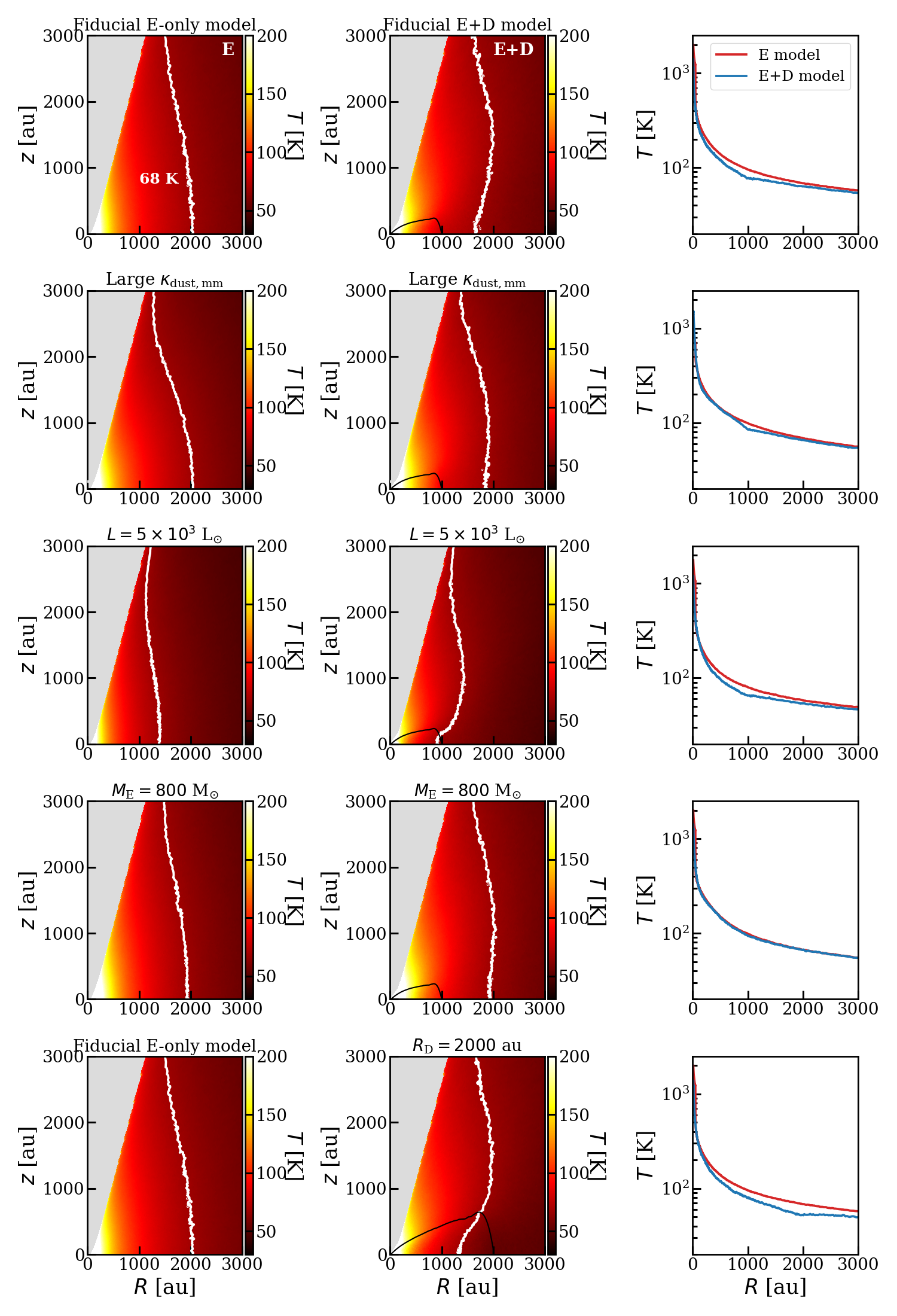}
    \caption{Temperature structure of models with different parameters. The left column presents the envelope-only models, the middle column shows the envelope-plus-disk models and the right column presents the comparison of a temperature cut at $z=0$\,au between the two models. The rows from top to bottom are for the fiducial model, that with large mm opacity dust, with protostellar luminosity of $5\times 10^3$\,L$_{\odot}$, with envelope mass of 800\,M$_{\odot}$ and finally the fiducial model with disk radius of 2000\,au.} 
    \label{fig:temp_grid}
\end{figure*}
\begin{figure}
  \resizebox{\hsize}{!}{\includegraphics{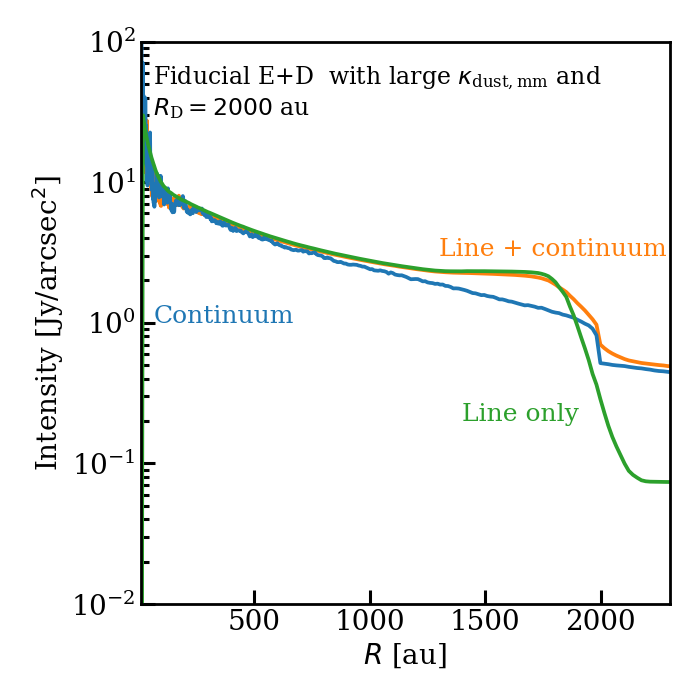}}
  \caption{An intensity cut through the fiducial model with large mm opacity dust and disk radius of 2000\,au. Orange presents when dust and gas are included in the run, blue presents when only dust is included and green presents when there is no dust included. We see that the continuum intensity is as large as the continuum plus line intensity in the inner ${\sim}1500$\,au and continuum subtraction will result in almost zero intensities. While, the intensity of the line-only run is as large as the line plus continuum run in the inner ${\sim}1500$\,au.}
  \label{fig:cont_over_sub_2000au}
\end{figure}
\begin{figure}
  \resizebox{\hsize}{!}{\includegraphics{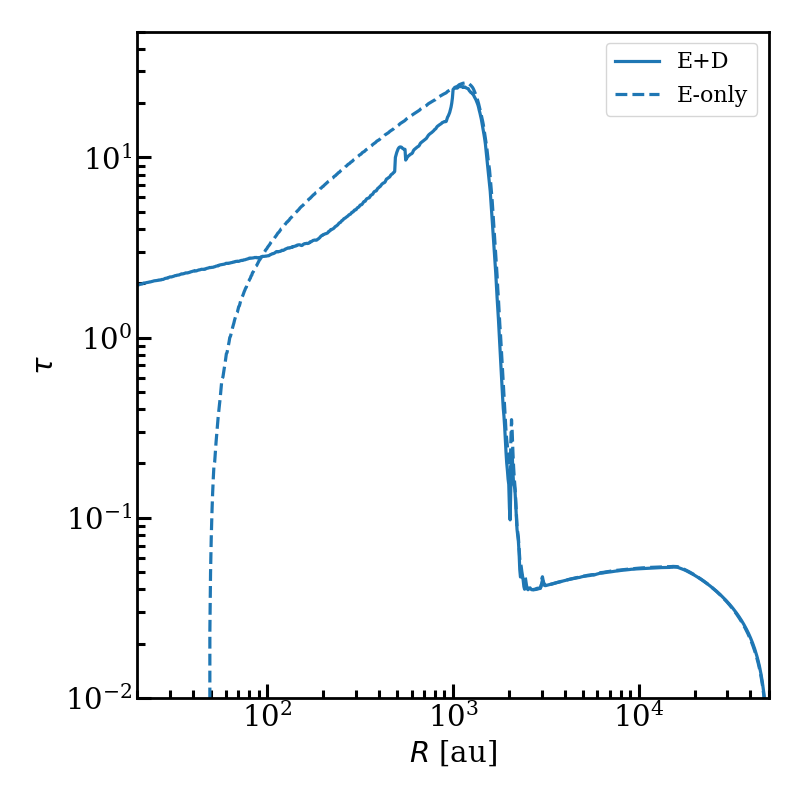}}
  \centering
  \caption{Methanol optical depth as a function of radius at the peak of the line. The dashed and solid lines show the fiducial envelope-only and envelope-plus-disk models, respectively. The emission is optically thick inside the snow surface.}
  \label{fig:tau_line}
\end{figure}
\begin{figure}
  \resizebox{\hsize}{!}{\includegraphics{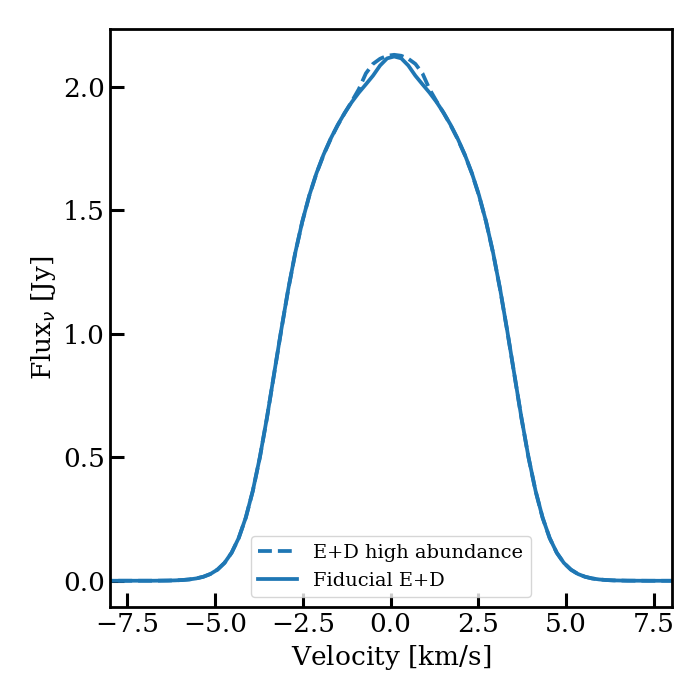}}
  \centering
  \caption{Methanol line emission for the fiducial envelope-plus-disk model (solid line) and the same with 2 orders of magnitude higher disk abundances (dashed line).}
  \label{fig:high_abund}
\end{figure}
\begin{SCfigure*}
    \centering
    \includegraphics[width=12cm]{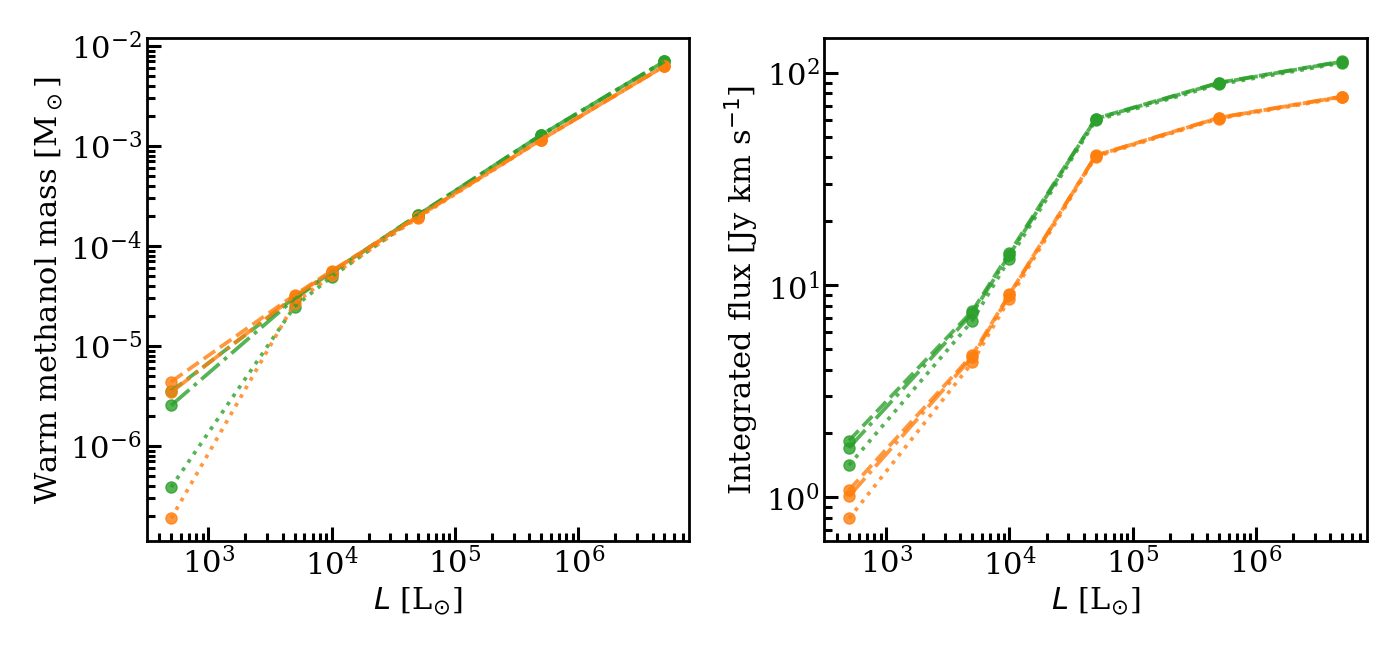}
    \caption{The same as Fig. \ref{fig:HII} but dashed lines show when the size of carved region is 50\,au, dashed dotted lines show the same for 200\,au and the dotted lines show the same for 500\,au.} 
    \label{fig:HII_small}
\end{SCfigure*}

Figure \ref{fig:temp_vertical_L5e5} is the same as Fig. \ref{fig:temp_vertical} but for a bolometric luminosity of $5 \times 10^5$\,L$_{\odot}$ and a vertical temperature cut at 30\,au. Figure \ref{fig:meth_emission} presents the methanol emission at the peak of the line viewed edge-on for the fiducial models. Moreover, this figure shows the line flux for the fiducial models. Figure \ref{fig:angle} presents the effect of viewing angle on the integrated flux of the fiducial envelope-only and envelope-plus-disk models and those with high mm opacity dust. Figures \ref{fig:meth_abund_M50} and \ref{fig:meth_abund_M150} are the same as Fig. \ref{fig:meth_abund} but for envelope masses of 50\,M$_{\odot}$ and 150\,M$_{\odot}$ respectively. Figure \ref{fig:temp_grid} presents the temperature structure of envelope-only and envelope-plus-disk models for various parameters varied in this work. Figure \ref{fig:cont_over_sub_2000au} shows that how continuum subtraction results in an error in the measured intensity for the fiducial model with large mm opacity and disk radius of 2000\,au. It particularly shows that the intensity of the continuum, line plus continuum and line-only runs are all the same in the inner ${\sim}1500$\,au. Figure \ref{fig:tau_line} presents the methanol line optical depth as a function of radius for the fiducial envelope-only and envelope-plus-disk models viewed face-on. Figure \ref{fig:high_abund} presents the methanol emission line for the fiducial envelope-plus-disk model and that with two orders of magnitude higher disk methanol abundances. Figure \ref{fig:HII_small} presents the warm methanol mass and integrated methanol flux for models with a simulated HII region for the envelope-only models by setting the methanol abundance to zero in an inner sphere. The radii assumed for the inner sphere in Fig. \ref{fig:HII_small} are 50\,au, 200\,au and 500\,au.


\end{appendix}
\end{document}